\documentclass[prb,reprint]{revtex4-2}
\usepackage{graphicx,amssymb,amsfonts,amsmath}
\usepackage{color}
\usepackage{mathtools}
\usepackage{latexsym}
\usepackage{graphicx}
\usepackage{float}
\usepackage{dcolumn}
\usepackage{bm}
\usepackage{amssymb}
\usepackage{amsmath}
\usepackage{mathtools}
\usepackage[space]{grffile}
\usepackage{xcolor}
\usepackage{comment}
\usepackage{hyperref}
\hypersetup{
    colorlinks=true,
    linkcolor=blue,
    filecolor=blue,      
    urlcolor=blue,
    citecolor=blue
    }

\begin{document}

\title{Anderson Localization: A Floquet operator Krylov space perspective}
\author{Hsiu-Chung Yeh$^1$}
\author{Aditi Mitra$^2$}
\affiliation{
$^1$Max Planck Institute for the Physics of Complex Systems, 01187 Dresden, Germany\\
$^2$Center for Quantum Phenomena, Department of Physics,
New York University, 726 Broadway, New York, New York, 10003, USA
}

\begin{abstract}
The problem of Anderson localization, as well as the single particle localization-delocalizaton transition of the Aubry-Andr\'e model, is studied employing operator Krylov space methods. It is shown that even when the dynamics is generated by a Hamiltonian,  studying the dynamics at stroboscopic rather than continuous times has its advantages. In particular, mapping the dynamics to an effective Floquet problem  results in an operator Krylov space description where quantities such as the spectral function can be computed with fewer computational resources, while a moment method exists that allows for the extraction of Krylov parameters directly from the discrete time autocorrelation function. For stroboscopic dynamics, the operator Krylov space corresponds to the dynamics of an edge operator of an 
inhomogeneous Floquet transverse field Ising model, with the parameters of this effective model generated recursively. The Krylov parameters show disorder-averaged renormalization with their distribution narrowing as the recursion step increases.
It is shown that a more physical spectral function is obtained from the Krylov parameters obtained from the disorder-averaged autocorrelation function, rather than the disorder-averaged Krylov parameters. 
The delocalized (localized) phase is shown to correspond to the appearance (absence) of a Porter-Thomas distribution, a ballistically propagating (localized) wavefront in operator Krylov space, and a smooth (discrete) Berstein-Szeg\"o power-spectrum. The localization-delocalization transition is also demonstrated in operator Krylov space. A Porter-Thomas distribution 
is also observed at the critical point. 
The long-time dynamics and the inverse participation ratio at the critical point is shown to exhibit behavior consistent with a multi-fractal scaling with system size.
\end{abstract}
\maketitle

\section{Introduction}

Operator Krylov space methods have emerged as a powerful approach to studying operator spreading in a variety of quantum many-body systems \cite{Recbook,parker2019universal,nandy2024quantum}. These include energy conserving dynamics generated by a Hamiltonian \cite{rabinovici2021operator,Yates20a,Yates20,yeh2023slowly}, stroboscopic dynamics generated by a unitary \cite{Yates21,yeh2023universal,yeh2025moment,yeh2026OPUC,suchsland2023krylov,kolganov2025streamlinedkrylovconstructionclassification}, and dissipative dynamics generated by a Lindbladian \cite{bhattacharya2022operator,bhattacharya2023krylov,bhattacharya2023krylov,liu2023krylov}.
All these methods revolve around recursively constructing a basis of operators, represented as sites of a Krylov chain. The length of the Krylov chain is typically exponentially large in the system size for many-body systems. 

The power of this method lies in the fact that in the Krylov basis, the operator dynamics can have a sparse and local representation, often with a  single-particle interpretation. The locality allows one to make physically motivated approximations for the "bulk" of the Krylov chain so as to capture the behavior in the thermodynamic limit. For example, when the dynamics is generated by a Hamiltonian, the Liouvillian superoperator is tri-diagonal in the Krylov basis, with the operator dynamics being mapped exactly to a single particle hopping between nearest-neighbor sites. The  hopping amplitudes on the chain are "spatially" inhomogeneous Lanczos coefficients that have certain universal features \cite{parker2019universal}, allowing for the development of coarse grained approaches for studying the dynamics \cite{Yates20,yeh2023slowly}. 

When the dynamics is generated by a unitary $U$, one is studying operator spreading at discrete or stroboscopic times.
In this case, it was shown that there exists a sparse and local operator Krylov space representation which is five-diagonal (reducing to the tri-diagonal form discussed above when the unitary is also Hermitian) \cite{CANTERO200329,CANTERO200540,SIMON2007120}. This basis has an elegant single particle interpretation in terms of a transverse field Ising model. In particular, the exact same operator dynamics can be generated by the edge operator of a Floquet transverse field Ising model with spatially inhomogeneous couplings (ITFIM), with the seed operator, i.e., the first site of the Krylov chain being the edge Majorana operator of the ITFIM \cite{yeh2023universal,yeh2025moment,yeh2026OPUC}.  

For a Hamiltonian $H$, the dynamics is generated by the unitary $e^{-i H t}$, where $t$ is continuous time. In this work we highlight the advantage of studying the dynamics not at arbitrarily closely separated times, but at discrete times $t = n T$, with $T$ being an artificially set period chosen to be not too short, and $n\in$ integer.  Thus we replace the continuous time dynamics by  stroboscopic Floquet dynamics generated by the unitary $U = e^{ -i H T}$. This mapping is then used to study longer times than would have been possible with the continuous time Krylov space approach.
We apply this method to the celebrated Anderson localization problem. 

The paper is organized as follows. In section \ref{FKrylov} we review the operator Krylov space methods for stroboscopic dynamics. In section \ref{Models} we provide the motivation for studying continuous time dynamics stroboscopically and present the Hamiltonians that will be studied. In section \ref{AndModel} we present results for the disordered one-dimensional (1D) Anderson model. In section \ref{AAmodel} we present results for the AA model, presenting results both in the localized phase, the delocalized phase, and at the critical point separating the two phases. In section \ref{Concl} we present our conclusions. Many details are relegated to appendices. 

\section{Floquet Operator Krylov Space} \label{FKrylov}
\label{Sec: Floquet Operator Krylov Space}

Hamiltonian dynamics is continuous time, energy conserving dynamics. However, on introducing an artificial discrete time $T$, the same dynamics can be converted to stroboscopic dynamics generated by a unitary $U=e^{-i H T}$. 
Operator spreading under any unitary $U$ has a compact Krylov space description in terms of a 1D kicked Ising model, dubbed the 1D Floquet ITFIM \cite{yeh2023universal,yeh2025moment,yeh2026OPUC}. We summarize these results  below.

We first outline the Arnoldi iteration for generating a Krylov basis. 
Given a unitary $U$, and a seed operator $O_0$, the repeated application of $(U^{\dagger})^nO_0 U^n=O_n$, generates new operators $O_n$. Denoting these operators as states $O_n\rightarrow |O_n)$, the next step is to Gram-Schmidt orthogonalize the states. At this stage, one needs to define an inner-product, and the natural one is the "infinite-temperature" trace
\begin{align}
     (A|B) = \frac{1}{\text{Tr}[\mathbb{I}]}\text{Tr}[A^\dagger B] - \frac{1}{\text{Tr}[\mathbb{I}]^2}\text{Tr}[A^\dagger]\text{Tr}[B],
     \label{Eq: inner prod}
\end{align}
where $\mathbb{I}$ denotes the identity operator. The second term, is included to simplify the resulting expressions for bilinear complex fermionic operators, which are the focus of this work (see Appendix \ref{App: A} for details). This is the essence of the Arnoldi iteration, with the unitary $U$ in the Krylov orthonormal basis  $\{ |\mathcal{O}_0), |\mathcal{O}_1), \ldots, |\mathcal{O}_{N-1}) \}$ having an upper-Hessenberg form \cite{Yates21,suchsland2023krylov,yeh2023universal} form. In particular, defining $\mathcal{K}_{ij}= (\mathcal{O}_i|U|\mathcal{O}_j)$, the explicit form for $\mathcal{K}$ for $N=6$ is
\begin{align}
    \mathcal{K}=\begin{pmatrix}
        a_0 & c_1 & c_2 & c_3 & c_4 & c_5\\
        b_1 & a_1 & \frac{a_1}{c_1}c_2 & \frac{a_1}{c_1}c_3 & \frac{a_1}{c_1}c_4 & \frac{a_1}{c_1}c_5\\
        0 & b_2 & a_2 & \frac{a_2}{c_2}c_3 & \frac{a_2}{c_2}c_4 & \frac{a_2}{c_2}c_5\\
        0 & 0 & b_3 & a_3 & \frac{a_3}{c_3}c_4 & \frac{a_3}{c_3}c_5\\
        0 & 0 & 0 & b_4 & a_4 & \frac{a_4}{c_4}c_5\\
        0 & 0 & 0 & 0 & b_5 & a_5
    \end{pmatrix}.\label{Eq: K matrix Krylov basis K6}
\end{align}
Above, $a_j, b_j, c_j$ have the following dependence on angles $\theta_j$ (dubbed the Krylov angles) \cite{yeh2023universal}
\begin{subequations} \label{Eq: K-matrix Krylov space2}
\begin{align}
    &a_j = \cos\theta_j\cos\theta_{j+1};\quad\quad\quad
    b_j = \sin\theta_j;\\
    &c_j = (-1)^j\cos\theta_{j+1} \prod_{k=1}^j\sin\theta_k,
    \end{align}
\end{subequations}
with $\theta_0 = \theta_{N} = 0$ and $\theta_i \in [0,\pi]$.

The upper-Hessenberg form is not a sparse representation, and also does not have a simple single-particle interpretation. Remarkably, another basis exists \cite{yeh2023universal} where the Krylov matrix takes a sparser, 5-diagonal form. For the $N=6$ example above, it is
\begin{widetext}
\begin{equation}
K_F=\begin{pmatrix}
\cos\theta_1 & \sin\theta_1 & 0 & 0 & 0 & 0 \\
-\sin\theta_1\cos\theta_2 & \cos\theta_1\cos\theta_2 & \sin\theta_2\cos\theta_3 & \sin\theta_2\sin\theta_3 & 0 & 0 \\
\sin\theta_1\sin\theta_2 & -\cos\theta_1\sin\theta_2 & \cos\theta_2\cos\theta_3 & \cos\theta_2\sin\theta_3 & 0 & 0 \\
0 & 0 & -\sin\theta_3\cos\theta_4 & \cos\theta_3\cos\theta_4 & \sin\theta_4\cos\theta_5 & \sin\theta_4\sin\theta_5 \\
0 & 0 & \sin\theta_3\sin\theta_4 & -\cos\theta_3\sin\theta_4 & \cos\theta_4\cos\theta_5 & \cos\theta_4\sin\theta_5 \\
0 & 0 & 0 & 0 & -\sin\theta_5 & \cos\theta_5
\end{pmatrix}.\label{K-5d}
\end{equation}
\end{widetext}
The same upper-Hessenberg matrix $\mathcal{K}$ is generated by the 1D Floquet ITFIM $U_F$, parametrized by identical Krylov angles and with open boundary conditions,
\begin{subequations}
\label{Eq: ITFIM}
\begin{align}
&U_{F} = U_z U_{xx};\\
&U_z = \prod_{l=1}^{N/2} e^{-i\frac{\theta_{2l-1}}{2}\sigma_l^z} ;
\ U_{xx} = \prod_{l=1}^{(N-2)/2} e^{-i\frac{\theta_{2l}}{2}\sigma_l^x \sigma_{l+1}^x}.
\end{align}  
\end{subequations}
Above, $\sigma^{x,z}_l$ are the Pauli  matrices on site $l$,
and the seed operator $ |O_0) =\sigma^x_1$. We assume $N$ is even  and $\{ \theta_1, \ldots, \theta_{N-1} \} \in [0,\pi]$ are the Krylov angles. The odd Krylov angles $\theta_{2l-1}$ denote the strength of the transverse field on site $l$ while the even Krylov angles $\theta_{2l}$ denote the strength of the Ising interaction between the spins on sites $l,l+1$. 
The ITFIM $U_F$ is a single-particle unitary as one can perform a Jordan-Wigner transformation and represent it as a binary drive involving Majorana fermion bi-linears. It is in the Majorana representation that the Krylov matrix has the five-diagonal form $K_F$, while in the spin-basis it has the upper-Hessenberg form $\mathcal{K}$. Thus, in some sense the basis transformation from the upper-Hessenberg to the five-diagonal form, derived originally using orthogonal polynomials \cite{CANTERO200329,CANTERO200540,SIMON2007120}, is equivalent to a Jordan-Wigner transformation of the 1D Krylov chain.

The mapping to a single-particle problem gives us certain insights on the dynamics, many of which come from the topological interpretation of the kicked or Floquet Transverse Field Ising model with uniform couplings. 
Consider, when all the even $\theta_{2l}$ (odd $\theta_{2l-1}$) Krylov angles are uniform 
and equal to $\theta_{\rm even} (\theta_{\rm odd})$. The phase diagram of this model is shown in the lower right panel of  Fig.~\ref{Fig: Anderson-1}. Depending on the parameters, the single-particle spectrum has gaps at zero and $\pi$ quasi-energies, with either no edge-modes (trivial) or $0,\pi$ edge modes. These edge modes represent a seed operator whose autocorrelation function is long lived. Thus the mapping immediately indicates that when the dynamics is chaotic, there should be no gap in the spectral density. This implies that $\theta_{\rm even}=\theta_{\rm odd}=\pi/2$ (center of the phase-diagram). 
Having $\theta_{\rm even}=\theta_{\rm odd}$ or $\theta_{\rm even}=\pi-\theta_{\rm odd})$ (the two diagonals in
the phase diagram) is not sufficient as it closes the gap at zero quasi-energy, while there is a gap at $\pi$
quasi-energy, or vice versa. 

We will see that both in the delocalized and the localized phases, since there is no gap in the disorder-averaged spectral density, the recursively generated Krylov angles will approach $\pi/2$. However, it is the manner in which this steady-state is reached, that will determine whether one is in the localized phase or the delocalized phase. 

We now discuss the numerical algorithms used to construct the ITFIM
for a given seed operator $O_0$ and a unitary $U$. There are primarily two methods, one employs Gram-Schmidt procedure, and the other employs a more efficient approach, which is dubbed the moment method \cite{yeh2025moment}. 

\subsection{Computing $\theta_j$ from Gram-Schmidt procedure} 
\label{Sec: Gram-Schmidt}
Gram-Schmidt orthogonalization is the most direct, but also the most computationally expensive way to determine the Krylov angles, i.e, the parameters of the ITFIM. The steps involve constructing matrix representations of the unitary, the seed operator, and then performing matrix multiplications, followed by the Gram-Schmidt orthogonalization. 

Based on the numerical results for $a_j, b_j, c_j$, the Krylov angles can be solved iteratively and uniquely. There are two possible iterative algorithms.  The first algorithm is 
\begin{subequations}
\begin{align}
    &\cos\theta_1 = a_0,\ \cos\theta_2 = \frac{a_1}{a_0}\\
    &\cos\theta_{j} = \frac{a_{j-1}}{a_{j-2}}\cos\theta_{j-2},\ \forall\ j\geq3.
\end{align}
\end{subequations}
Note that the above relations are numerically unstable when  the prefactor is the ratio of small numbers, namely $a_j \rightarrow 0\ (\theta_j \rightarrow \pi/2)$  for large $j$.

The second algorithm is the following
\begin{align}
    \cos\theta_1= a_0,\ \cos\theta_{j+1} = \frac{(-1)^{j}c_j}{b_j b_{j-1} \ldots b_1}.
\end{align}
This is numerically unstable when the denominator is small, namely $b_j b_{j-1} \ldots b_1 \rightarrow 0$ for large $j$. Therefore, this algorithm is stable if $\theta_j \rightarrow \pi/2$ fast enough.

Depending on the asymptotic behavior of the Krylov angles, one can adopt one of the above algorithms accordingly. 
For chaotic models as well as the single-particle Anderson model, one expects no gap in the spectral density, and we will find setting $\theta_{k>k^*} =\pi/2$, after a certain number of recursions $k^*$, is an efficient way to approximate a system in the thermodynamic limit. 

The most time consuming part of the numerical computation is the Gram-Schmidt orthogonalization. We now discuss a second algorithm, the moment method, that allows us to completely avoid the Gram-Schmidt procedure. 

\subsection{Computing $\theta_j$ from the Moment Method}
\label{Sec: moment method}

The key observation of the moment method is that an
autocorrelation of the seed operator can be related to the matrices in the Krylov  representation as follows
\begin{align}
    A(nT) = (O_0| (U^\dagger)^n O_0 U^n) = (\mathcal{O}_0|K^n_F|\mathcal{O}_0). \label{Eq: autocorrelation}
\end{align}
Note that the unitary has stroboscopic time dependence; therefore, we write $A(nT)$ to make this dependence explicit.
{Substituting the explicit form of $K_F$, \eqref{K-5d}, allows one to solve for the Krylov angles from $A(nT)$ as follows
\begin{subequations}
\label{Eq: Angles and Autocorrelation}
\begin{align}
    &A(T) = \cos\theta_1, \label{Eq: A(1)}\\
    &A(2T) = \cos^2\theta_1 - \sin^2\theta_1 \times \cos\theta_2.\label{Eq: A(2)}
\end{align}
\end{subequations}
In general, $A(nT)$ has the following form
\begin{align}
    A(nT) = f(n-1) + (-1)^{n-1}\cos\theta_n \prod_{k=1}^{n-1} \sin^2\theta_k, \label{Eq: Autocorrelation to angle}
\end{align}
where $f(n-1)$ is some function that depends on the Krylov angles $\{\theta_1, \theta_2, \ldots, \theta_{n-1}\}$, with the dependence on $\theta_n$ of $A(nT)$ only appearing in the second term.

Numerically, the Krylov angles are solved order by order according to
\begin{align}
    \cos\theta_n = \frac{A(nT)-f(n-1)}{(-1)^{n-1}\prod_{k=1}^{n-1} \sin^2\theta_k}.
    \label{Eq: moment algo}
\end{align}
In general, the moment method is more efficient in computing Krylov angles as only the autocorrelation is computed instead of the full Krylov space orthogonal basis.  This numerical algorithm might fail when $\prod_{k=1}^{n-1} \sin^2\theta_k$ becomes too small, which is also the drawback of the second algorithm presented in the previous subsection. However, for models where $\theta_j\rightarrow \pi/2$ fairly rapidly, the moment method is a stable numerical method. 

\subsection{The Bernstein–Szeg\H{o} approximation for the spectral function}

Any operator can be expanded in the orthogonal basis of operators generated by the Krylov method. This basis also has a convenient representation in the form of orthogonal polynomials on the unit circle (OPUC) \cite{simon2005orthogonalpart1}. These polynomials have many interesting properties allowing one to efficiently compute physical quantities such as spectral functions. We highlight this application below.

After solving for the Krylov angles numerically, one can first construct the OPUC, $P_k(z)$, according to the Szeg\H{o} recurrence relation \cite{yeh2026OPUC,simon2005orthogonalpart1}
\begin{align}
    &\begin{pmatrix}
        P_{k}(z)\\
        P_{k}^*(z)
    \end{pmatrix}\nonumber\\ 
    &=  \frac{1}{\sin\theta_k}
    \begin{pmatrix}
        z & (-1)^k \cos\theta_k \\
        (-1)^k \cos\theta_k \,z &  1
    \end{pmatrix}
    \begin{pmatrix}
        P_{k-1}(z)\\
        P_{k-1}^*(z)
    \end{pmatrix},
    \label{Eq: Szego recurrence}
\end{align}
where the initial condition is $(P_0(z), P_0^*(z)) = (1,1)$.
Note that $P_k^*$ is not complex conjugation but a $*$-reverse polynomial of $P_k$: $P_k^* = z^k \overline{P_k(1/\overline{z})}$. The complex conjugation of the complex number $z$ is denoted as $\overline{z}$.

The advantage in using OPUC is that one can readily compute approximate spectral functions. In particular, the Bernstein–Szeg\H{o} approximation for the spectral function when truncating at $k_*$ and setting $\theta_{k>k^*}=\pi/2$ is {\cite{simon2005orthogonalpart1}
\begin{align}
    \Phi_{k_*}(e^{i\omega}) = \frac{1}{2\pi}   \frac{1}{|P_{k_*}(e^{i\omega})|^2}.
    \label{Eq: Bernstein approx}
\end{align}
The exact spectral function is recovered by taking the following limit
\begin{align}
    \Phi(e^{i\omega}) = \frac{1}{2\pi} \lim_{k_{*}\rightarrow \infty}  \frac{1}{|P_{k_*}(e^{i\omega})|^2}.
    \label{Eq: Bernstein large k}
\end{align}

In order to appreciate the Bernstein–Szeg\H{o} approximation for the spectral function, let us compare it with the standard evaluation for the spectral function.
The spectral function is formally defined as the Fourier transform of the autocorrelation,
\begin{align}
    \Phi(e^{i\omega}) = \frac{1}{2\pi} \sum_{n=-\infty}^{\infty} A(nT)e^{i\omega n}.
    \label{Eq: Fourier}
\end{align}
In practical computations, this infinite sum must be truncated at some finite value $n_*$. Accordingly, the truncated Fourier transform of the spectral function is given by
\begin{align}
    \Phi^{(n_*)}(e^{i\omega}) = \frac{1}{2\pi}\sum_{n=-n_*}^{n_*}A(nT)e^{i\omega n}.
    \label{Eq: truncated Fourier}
\end{align}
Both the Bernstein--Szeg\H{o} approximation, $\Phi_{k_*}$, and the truncated Fourier transform, $\Phi^{(n_*)}$, are constructed from the same autocorrelation data when $k_* = n_*$. A subtlety of $\Phi^{(n_*)}$ is that artificial oscillations may appear in numerical implementations, as the cancellation from higher harmonics is lost in the truncation procedure of \eqref{Eq: truncated Fourier}. Since only $(2n_* + 1)$ values of $A(nT)$ are used, we compute the spectral density at $(2n_* + 1)$ independent frequencies with spacing $2\pi/(2n_* + 1)$. In practice, one would need to align these discrete frequencies with the intrinsic frequencies present in $A(nT)$ in order for $\Phi^{(n_*)}$ to behave smoothly when evaluated on this grid. However, this alignment is generally difficult to achieve when the autocorrelation contains multiple intrinsic frequencies and the available data are limited. 

In contrast, the Bernstein--Szeg\H{o} approximation can be interpreted as a soft truncation, as it effectively assumes $\theta_{k > k_*} = \pi/2$. This typically results in a smoother and more accurate approximation compared to the truncated Fourier transform.

\section{Models} \label{Models}

We now turn to applications. We will study Hamiltonian dynamics using Krylov recursion methods. The standard approach for continuous time dynamics is as follows. Given a seed operator $O_0$, its Heisenberg time-evolution is
\begin{align}
    O(t) = e^{i H t} O_0 e^{-i Ht} = \sum_{n=0}^{\infty}\frac{(it)^n}
    {n!}\mathcal{L}^n O_0, \mathcal{L} = \left[H,\cdot\right].
\end{align}
The Krylov recursion method used in this context, generates a basis where $\mathcal{L}$  is tri-diagonal in the Lanczos coefficients $\Tilde{b}_n$,
\begin{align}
    \mathcal{L} = \begin{pmatrix}
        0 & \Tilde{b}_1 &\\
        \Tilde{b}_1 & 0 & \Tilde{b}_2\\
        & \Tilde{b}_2 & 0 & \Tilde{b}_3 \\
        && \ddots & \ddots & \ddots
    \end{pmatrix}.
\end{align}
Despite this simplicity, there is no direct relation between the first $n$ Lanczos coefficients and the time up to which one can simulate the dynamics.  To see this explicitly, consider the continuous time autocorrelation function 
\begin{align}
    C(t) = (O_0|O_0(t)).
\end{align}
The first $n$ moments of the autocorrelation function, defined as
\begin{align}
m_n =  \lim_{t\to 0} \frac{d^nC(t)}{i^n dt^n} = 
(O_0|\mathcal{L}^n|O_0),
\label{Eq: moments C}
\end{align}
do depend only on  the first $n$ Lanczos coefficients,
\begin{subequations}
\label{Eq: Lanczos moments}
\begin{align}
    &m_2 = \Tilde{b}_1^2,\\
    &m_4 = \Tilde{b}_1^2 \Tilde{b}_2^2 + \Tilde{b}_1^4,\\
    &m_{2n} = g(n-1) + \Tilde{b}_1^2 \Tilde{b}_2^2 \ldots \Tilde{b}_{n}^2,
\end{align}
\end{subequations}
where $g(n-1)$ is some function that depends on $\{\Tilde{b}_1, \Tilde{b}_2, \ldots, \Tilde{b}_{n-1}\}$. However, constructing $C(t)$ from the moments requires knowing all the moments because $C(t) = \sum_{n=0}^\infty m_n(it)^n/(n!)$. Thus, arbitrarily short times require the knowledge of all Lanczos coefficients. 

In contrast, when time is discrete, i.e., the time-evolution is studied stroboscopically, the first $n$ Krylov angles, generate the dynamics upto the first $n$ stroboscopic times.
In addition, employing the moment method, one can completely bypass the Gram-Schmidt orthogonalization procedure, and use directly the knowledge of $A(nT)$ alone to generate the Krylov angles. Numerically, it is usually more efficient to generate an autocorrelation function, than to generate a Krylov basis upto long times. 

In what follows, we will study a disordered, non-interacting fermionic Hamiltonian at stroboscopic times $nT$, where $T$ is our chosen discretization time-step, and $n$ is the discrete time index. 
The Hamiltonians  of interest are of the form
\begin{align}
    H = \sum_{j=1}^{L-1}\biggl[c_j^\dagger c_{j+1} + c_{j+1}^\dagger c_j\biggr] + \sum_{j=1}^{L}h_j c_j^\dagger c_j.
\end{align}
In the bilinear complex fermionic basis, the Hamiltonian can be written as
\begin{align}
    H = \sum_{jk}\Tilde{H}_{jk} c_j^\dagger c_k,
\end{align}
where $\Tilde{H}$ is the following tri-diagonal matrix 
\begin{align}
    \Tilde{H} = \begin{pmatrix}
        h_1 & 1 &\\
        1 & h_2 & 1\\
        & 1 & h_3 & 1 \\
        && \ddots & \ddots & \ddots
    \end{pmatrix}.
\end{align}
The autocorrelation function~\eqref{Eq: autocorrelation} for bilinear fermionic operators can be written as (see Appendix \ref{App: A} for the derivation)
\begin{align}
    A(nT) = \frac{1}{4}\text{tr}\!\left[\Tilde{O}_0 (\Tilde{U}^\dagger)^n \Tilde{O}_0 \Tilde{U}^n \right], \label{Eq: auto bilinear}
\end{align}
where $\text{tr}[\cdot]$ denotes the trace taken in the bilinear complex fermionic basis, in contrast to $\text{Tr}[\cdot]$ that denotes the trace in the many-body Hilbert space. The matrices $\Tilde{O}_0$ and $\Tilde{U}$ are defined in the bilinear complex fermionic basis as
\begin{align}
     O_0 = \sum_{j,k} \Tilde{O} _{0jk} \, c_j^\dagger c_k, \quad \Tilde{U} = e^{-i \Tilde{H} T}.
\end{align}
In this work, we consider an operator of the form
\begin{align}
    \Tilde{O}_{0jk} = 2\delta_{j,\frac{L}{2} + 1}\delta_{k,\frac{L}{2} + 1},
\end{align}
which has a single nonzero diagonal element in the bilinear complex fermionic basis. The pre-factor of $2$ is chosen so that $A(0)=1$. For this choice of operator, the autocorrelation~\eqref{Eq: auto bilinear} can be interpreted as the return probability to the site $(1+L/2)$,
\begin{align}
    A(nT) = |\langle \phi(nT) | \phi \rangle|^2,
    \label{Eq: A return}
\end{align}
where $|\phi\rangle = (0,\ldots,0,1,0,\ldots,0)^\intercal$ has a nonzero component only at the $(1+L/2)$-th site, and $|\phi(nT)\rangle = \Tilde{U}^n |\phi\rangle$. Consequently, the long-time behavior of the autocorrelation serves as a probe of the system's phase: a non-decaying autocorrelation indicates localization.

To probe the late-time dynamics, we choose the stroboscopic time step $T$ in the numerical results sufficiently large to access the asymptotic regime of the autocorrelation.

\begin{figure*}
    \raggedright
    \includegraphics[width=0.32\textwidth]{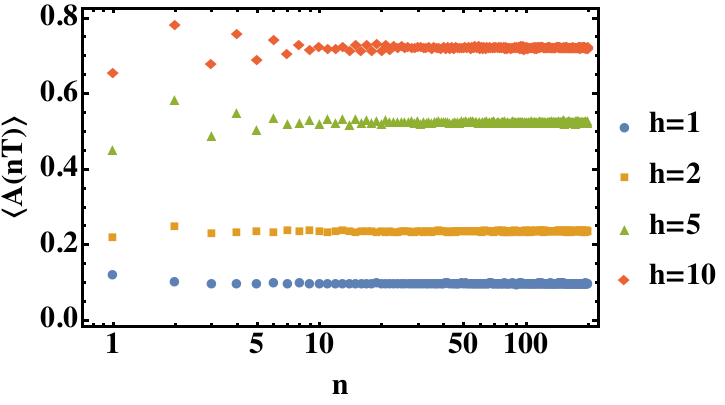}
    \includegraphics[width=0.32\textwidth]{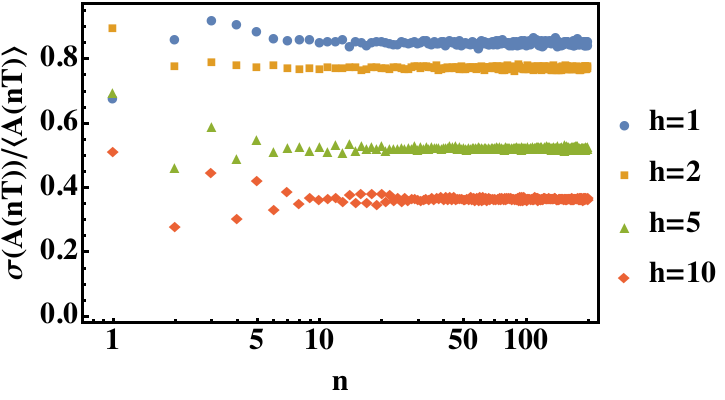}
    \includegraphics[width=0.32\textwidth]{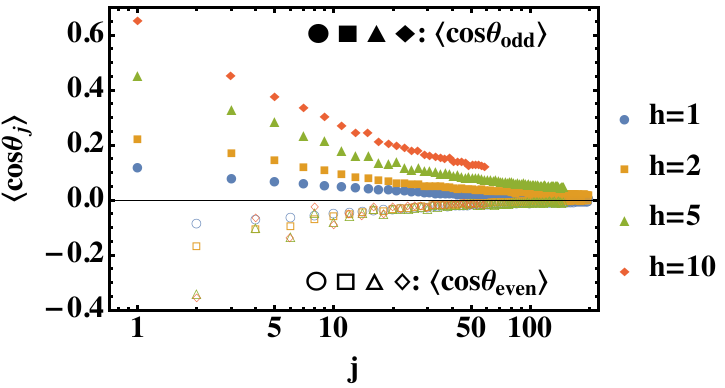}
    \includegraphics[width=0.32\textwidth]{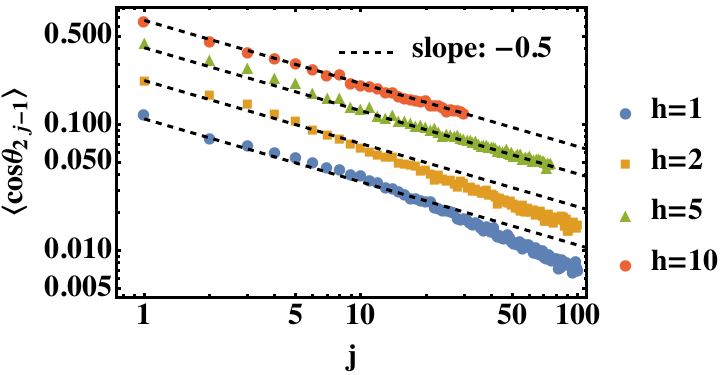}
    \includegraphics[width=0.32\textwidth]{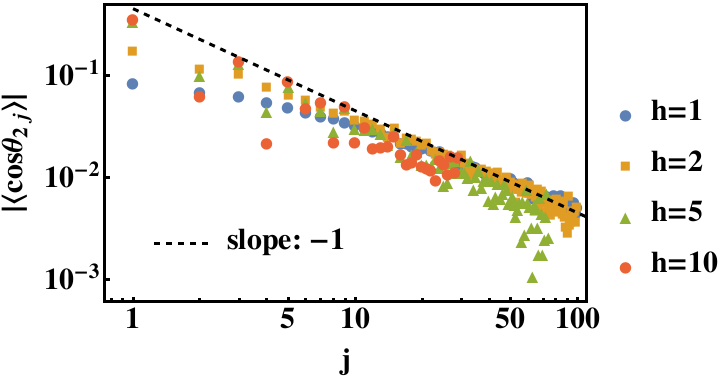}
    \includegraphics[width=0.25\textwidth]{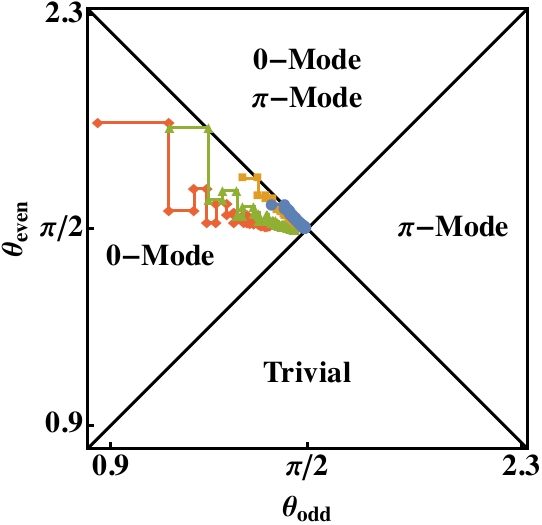}

    \caption{Results for the Anderson model for $L = 200$ and $T = 1.5$, with disorder strengths $h = 1, 2, 5, 10$, averaged over $15{,}000$ realizations. Top panels ($x$-axes are on a log-scale): the disorder-averaged autocorrelation saturates to a constant value at late times (left panel). The relative standard deviation of the autocorrelation also saturates at late times (middle panel). The disorder averaged cosine of the Krylov angles is shown in the right panel, where the odd/even angles approach $\pi/2$ from above/below. Bottom panels: log--log plots of the averaged cosines of the odd (left panel) and even (middle panel) Krylov angles. For strong disorder, $\langle \cos\theta_{2j-1} \rangle \sim 1/\sqrt{j}$, while $|\langle \cos\theta_{2j} \rangle| \sim 1/j$ for all disorder strengths. The rightmost panel shows the trajectory of the averaged Krylov angles in the phase diagram of kicked Ising model, where the odd (even) angles correspond to transverse fields (Ising couplings).}
    \label{Fig: Anderson-1}
\end{figure*}

\begin{figure*}
    \centering
    \includegraphics[width=0.33\textwidth]{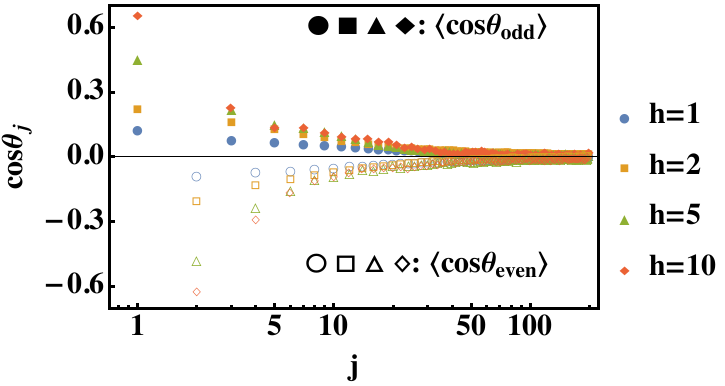} 
    \includegraphics[width=0.35\textwidth]{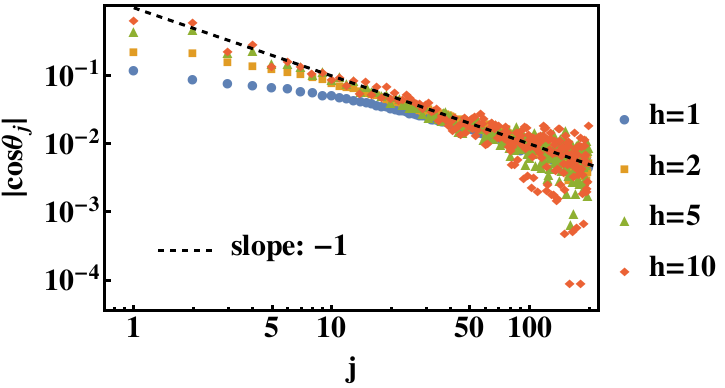}
    \includegraphics[width=0.25\textwidth]{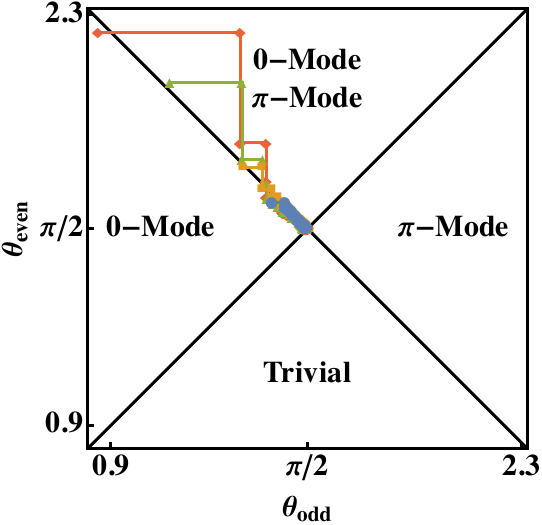}
    \caption{The Krylov angles are computed from the disorder-averaged autocorrelation (upper left panel of Fig.~\ref{Fig: Anderson-1}) and shown in the left panel. In contrast to the results in Fig.~\ref{Fig: Anderson-1}, both the odd and even Krylov angles exhibit the same asymptotic power-law behavior, $|\cos\theta_j| \sim 1/j$ for large $j$ (middle panel). Consequently, the trajectory of the Krylov angles in the phase diagram (right panel) lies along the diagonal. These results are consistent with the analytic solution for a constant autocorrelation which suggests a power-law localized $0$-mode in Krylov space~\cite{yeh2025moment}.}
    \label{Fig: Anderson-4}
\end{figure*}

\section{Anderson Model} \label{AndModel}
For the Anderson model, the local fields are drawn from a uniform random distribution,
\begin{align}
    h_j \in [-h,h].
\end{align}
We compute the autocorrelation for each disorder realization, and average it over disorder realizations. The notation $\langle A(nT)\rangle$ indicates the disorder averaged autocorrelation. In addition, we compute the standard deviation defined  as 
\begin{align}
    \sigma(A(nT)) = \sqrt{\langle A^2(nT)\rangle -  \langle A(nT)\rangle^2} \label{eq:var} 
\end{align}
From each disorder realization, a $A(nT)$ generates Krylov angles $\theta_j$. When these angles are averaged over different disorder realizations, we denote them as $\langle \cos\theta_j\rangle$.

In Fig.~\ref{Fig: Anderson-1}, we consider disorder strengths $h = 1, 2, 5, 10$ for a system of size $L = 200$, and a stroboscopic time step $T = 1.5$. For each disorder strength, the results are averaged over $15{,}000$ realizations. The disorder-averaged autocorrelation $\langle A(nT) \rangle$ (upper left panel of Fig.~\ref{Fig: Anderson-1}) saturates to a constant value, reflecting the fact that the 1D Anderson model is localized for any disorder strength. As expected, the relative standard deviation (upper middle panel of Fig.~\ref{Fig: Anderson-1}) decreases as the disorder strength increases.

The Krylov angles are computed for each realization using the moment method described in Sec.~\ref{Sec: moment method}. The disorder-averaged cosine of the Krylov angles, $\langle \cos \theta_j \rangle$ (upper right panel of Fig.~\ref{Fig: Anderson-1}), decays to zero for large $j$. Moreover, $\langle \cos \theta_{\text{odd/even}} \rangle$ approaches zero from above and below, respectively, exhibiting distinct power-law behaviors. For odd angles, $\langle \cos \theta_{2j-1} \rangle \sim 1/\sqrt{j}$ at strong disorder, while for weaker disorder a crossover behavior is observed. The cosine of the even angles decay faster, approximately as $ \langle \cos \theta_{2j} \rangle \sim -1/j$, for all disorder strengths. 

Although the autocorrelation exhibits a transient behavior at early times that depends on $T$, the Krylov angles shown in Fig.~\ref{Fig: Anderson-1}, at the same timesteps (Krylov indices) are less sensitive to $T$. This indicates, that the angles, for the same discrete time index are already capturing the late-time dynamics effectively. In Appendix \ref{App: Anderson numerical details}, we present additional results with $T = 30$ to further support this observation. Because the moment method becomes numerically unstable when $\cos\theta_j$ decays slowly, we terminate the iterations earlier in the strong-disorder regime. As a result, fewer data points are shown in Fig.~\ref{Fig: Anderson-1}, see discussion in Appendix \ref{App: Anderson numerical details} that compares the moment method with the numerically more costly Arnoldi iteration.

An alternative way to present the Krylov angles is to plot them in the phase diagram of the Floquet transverse-field Ising model, as shown in the lower right panel of Fig.~\ref{Fig: Anderson-1}. In this representation, the odd (even) Krylov angles correspond to Ising couplings (transverse fields). The trajectory of the Krylov angles is constructed from the sequence of points: $(\theta_1, \theta_2), (\theta_3, \theta_2), (\theta_3, \theta_4), \ldots$. Note that the angles are obtained from the disorder-averaged cosines via $\{\cos^{-1}(\langle \cos \theta_j \rangle)\}$. The resulting trajectories in the phase diagram indicate that the asymptotic behavior approaches the critical point diagonally for weak disorder and horizontally for strong disorder. This distinction arises from the slower decay of $\langle \cos \theta_{2j-1} \rangle$ relative to $\langle \cos \theta_{2j} \rangle$
for strong disorder.

The non-decaying behavior of the autocorrelation indicates that $O_0$ evolves into a conserved operator at late times. Since the trajectory in the phase diagram remains within the $0$-mode phase, this conserved operator can be interpreted as an edge $0$-mode in Krylov space. The corresponding $0$-mode wavefunction in Krylov space satisfies \cite{yeh2023universal}
\begin{align}
    \left|\frac{\psi_{2j+1}}{\psi_1}\right|^2 &= \prod_{k=1}^{j} \left[\tan\left(\frac{\theta_{2k-1}}{2}\right)\cot\left(\frac{\theta_{2k}}{2}\right)\right]^2\nonumber\\
    &= \prod_{k=1}^{j} \frac{1-\cos\theta_{2k-1}}{1+\cos\theta_{2k-1}}\frac{1+\cos\theta_{2k}}{1-\cos\theta_{2k}}.
    \label{Eq: wave function}
\end{align}
From numerical observations, the typical asymptotic behavior of the Krylov angles at large $j$ in the strong-disorder regime is
\begin{align}
    \langle\cos\theta_{2j-1}\rangle  \sim \frac{\eta_o}{\sqrt{j}},\ \langle\cos\theta_{2j}\rangle\sim -\frac{\eta_e}{j},\ \forall\ j \gg 1,
    \label{Eq: ave Krylov angles}
\end{align}
where $\eta_{o/e}$ is a constant that depends on the disorder strength. 

The typical asymptotic behavior of the Krylov angles observed above is closely related to the geometric average of the $0$-mode wavefunction. To make this connection explicit, we consider the geometric average over disorder realizations, labeled by $r$:
\begin{align}
    &\left(\prod_{r=1}^{R} \left|\frac{\psi_{2j+1,r}}{\psi_{1,r}}\right|^2\right)^{1/R}\nonumber\\
    &=  \prod_{r=1}^{R}\left(\prod_{k=1}^{j} \frac{1-\cos\theta_{2k-1,r}}{1+\cos\theta_{2k-1,r}}\frac{1+\cos\theta_{2k,r}}{1-\cos\theta_{2k,r}} \right)^{1/R}\nonumber\\
    &\sim \prod_{r=1}^{R}\left(e^{-2\sum_{k=1}^j (\cos\theta_{2k-1,r}-\cos\theta_{2k,r})}\right)^{1/R}\nonumber\\
    &\sim e^{-\frac{2}{R}\sum_{r=1}^R\sum_{k=1}^j (\cos\theta_{2k-1,r}-\cos\theta_{2k,r})}\nonumber\\
    &\sim e^{-2\sum_{k=1}^j (\langle\cos\theta_{2k-1}\rangle-\langle\cos\theta_{2k}\rangle)} \sim e^{-4\eta_o\sqrt{j}}
    \label{Eq: wave function geo ave}
\end{align}
where $R$ denotes the number of realizations. In the third line, we have assumed that the cosines of the Krylov angles are small, allowing for an exponential approximation. In the last line, the contribution from $\langle \cos\theta_{2k} \rangle$ is subleading and is therefore neglected. Interestingly, a stretched-exponential localization with a square-root exponent in the real space has been observed in random hopping models~\cite{inui1994unusual}. This result of stretched-exponential localization in the Krylov space is consistent with recent numerical studies~\cite{peacock2025anderson}, which focus on systems in higher dimensions.

Alternatively, one may first average the autocorrelation function over different realizations and then extract the Krylov angles from the averaged data. As we show below this averaging procedure is more mathematically well defined, giving physically sensible results for the spectral function. The results for this averaging procedure are presented in Fig.~\ref{Fig: Anderson-4}, which uses the same autocorrelation data as Fig.~\ref{Fig: Anderson-1}. In contrast to Fig.~\ref{Fig: Anderson-1}, the cosine of the Krylov angles decays as $1/j$ for both odd and even angles, and the trajectory in the phase diagram lies along the diagonal, for all disorder strengths. In addition, since the cosine of the Krylov angles decay more rapidly for this case, there is no need to terminate the iterations early even in the strong-disorder regime.

As discussed above, a non-decaying autocorrelation implies a 0-mode in Krylov space. 
The 0-mode obtained from the second procedure is power-law localized and is therefore quite different from the 0-mode from the previous procedure, which effectively corresponded to a geometric average of the 0-mode for each disorder realization. In the second approach, the resulting $0$-mode wavefunction in Krylov space is directly determined 
from the disorder averaged autocorrelation function, and we will argue below that this method leads to a more physical spectral function than the previous method. We also note that the result in Fig.~\ref{Fig: Anderson-4} can be related to the known analytic example of a completely constant autocorrelation, for which the Krylov angles do decay as $1/j$, and correspond to a $0$-mode wavefunction that is power-law localized~\cite{yeh2025moment}.

\begin{figure*}
    \centering
    \includegraphics[width=0.4\textwidth]{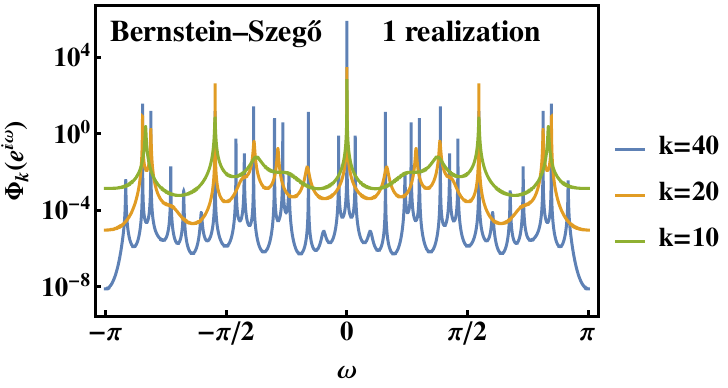} 
    \includegraphics[width=0.4\textwidth]{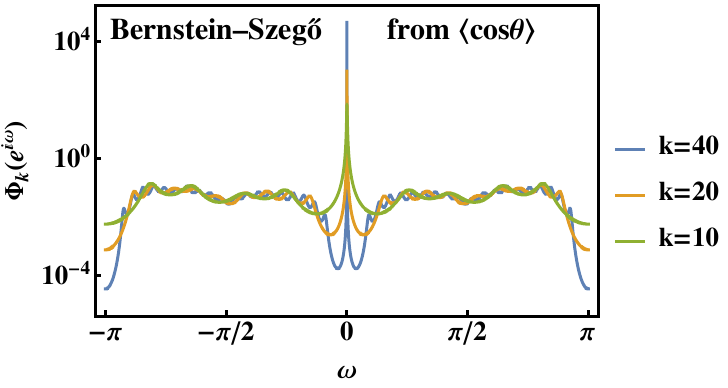}
    
    \includegraphics[width=0.4\textwidth]{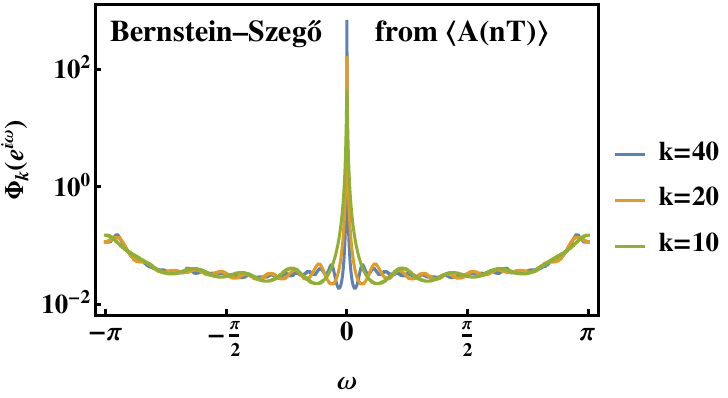} 
    \includegraphics[width=0.4\textwidth]{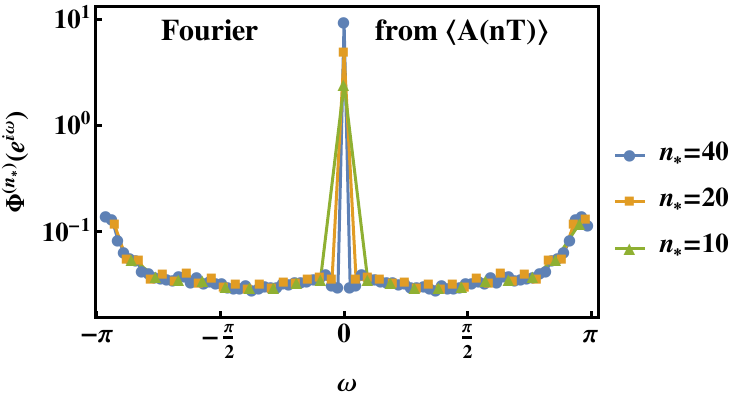}  

    \caption{Approximate spectral functions for the Anderson model obtained from different approximation schemes for $L=200,\ T=1.5,\ h=10$. Top panels: Bernstein--Szeg\H{o} approximation~\eqref{Eq: Bernstein approx} applied to a single realization (left panel) and to the average over $15{,}000$ realizations of the cosine of the Krylov angles (right panel). 
    Bottom panels: Approximate spectral function obtained from the Bernstein--Szeg\H{o} approximation~\eqref{Eq: Bernstein approx} based on Krylov angles extracted from the disorder-averaged autocorrelation (left panel) ; truncated Fourier transform~\eqref{Eq: truncated Fourier} computed from the disorder-averaged autocorrelation is evaluated only at $(2n_*+1)$ discrete values of the frequencies: $\omega = 2m\pi/(2n_*+1)$ with $m = -n_*, -n_*+1, \ldots, n_*$ (right panel).}
    \label{Fig: Anderson-5}
\end{figure*}

We now discuss the spectral functions obtained from the Bernstein--Szeg\H{o} approximation~\eqref{Eq: Bernstein approx} and the truncated Fourier transform~\eqref{Eq: truncated Fourier} in Fig.~\ref{Fig: Anderson-5}. Since the truncated Fourier transform is constructed from $(2n_* + 1)$ data points of the disorder-averaged autocorrelation and the corresponding spectral function is expected to be peaked at $\omega = 0$, it is natural to evaluate it only at the $(2n_* + 1)$ discrete frequency values $\omega = 2\pi m/(2n_* + 1)$, with $m = -n_*, -n_* + 1, \ldots, n_*$. The Bernstein--Szeg\H{o} approximation based on the Krylov angles from the two disorder-averaging schemes, one where cosine of the Krylov angles are averaged, and the other where the Krylov angles are extracted from the disorder averaged autocorrelation, yield a smoother reconstruction than the truncated Fourier transform. However, the two approaches exhibit different behavior near $\omega = 0, \pi$. The approximate spectral function obtained from disorder-averaged Krylov angles (upper right panel of Fig.~\ref{Fig: Anderson-5}) shows dips near $\omega = 0, \pi$, whereas the one obtained from the averaged autocorrelation (lower left panel) exhibits a comparatively flat spectrum in these regions. A discussion of the formation of the dips in the spectral function for the first averaging procedure is presented in Appendix~\ref{App: spectral dips}.

In Fig.~\ref{Fig: Spectral comparison}, we present numerical results obtained from both the Bernstein--Szeg\H{o} approximation ($k_* = 40$) and the truncated Fourier transform ($n_* = 40$), both obtained from the disorder-averaged autocorrelation.  We compare these two to the spectral function computed from the Lehmann representation, derived by combining \eqref{Eq: Fourier} and \eqref{Eq: A return},
\begin{align}
    \Phi(e^{i\omega}) = \sum_{j,k}|\langle \epsilon_j|\phi\rangle\langle\phi|\epsilon_k\rangle|\delta_F(\omega-\epsilon_j+\epsilon_k),
    \label{Eq: Lehmann}
\end{align}
where $U | \epsilon_k \rangle = e^{-i\epsilon_k} | \epsilon_k \rangle$, and $\delta_F(\omega)$ is the Floquet Dirac delta function defined as $\delta_F(\omega) = \sum_m \delta_D(\omega + 2\pi m)$, with $\delta_D(\omega)$ being the standard Dirac delta function. In numerical calculations, the Floquet delta function is approximated by a wrapped Lorentzian distribution,
\begin{align}
    \delta_F(\omega) \approx \frac{1}{2\pi}\frac{\sinh\eta}{\cosh\eta-\cos\omega},
    \label{Eq: wrapped Lorentzian}
\end{align}
where the width parameter $\eta$ is typically chosen to be of the order of the quasi-energy level spacing. The results shown in Fig.~\ref{Fig: Spectral comparison} are obtained by averaging over $5000$ realizations with $\eta = \pi/(2L)$ and $L = 200$. The dependence on $\eta$ and $L$ is discussed in Appendix~\ref{App: Lehmann}. All three methods show good overall agreement; however, the Bernstein--Szeg\H{o} approximation more efficiently captures the delta-function peak at $\omega = 0$. In contrast, the spectral function obtained from the disorder averaged Krylov angles show an unusual dip at low frequencies not exhibited by the disorder averaged Lehmann representation.  This strongly indicates that it is the second averaging scheme, where the autocorrelation is averaged first, is physically more reasonable.

We note that in the Hamiltonian Krylov method, obtaining Krylov parameters from the disorder averaged autocorrelation function is numerically more demanding. Although one can evaluate the continuous time auto-correlation and disorder-average it, it is never practical to use \eqref{Eq: moments C} directly as there is never enough data to perform a satisfactory numerical differentiation. 
In practice therefore, for each realization, one must first computes the Lanczos coefficients and then evaluate the corresponding moments according to \eqref{Eq: Lanczos moments}. The moments are then averaged over disorder realizations, after which the Lanczos coefficients corresponding to the averaged moments are extracted using \eqref{Eq: Lanczos moments}, again. In contrast, for the discrete time problem, the autocorrelation is directly related to the Krylov angles, as shown in \eqref{Eq: Angles and Autocorrelation}. As a result, the procedure is considerably more straightforward and numerically stable once one discretizes the Hamiltonian time-evolution.

An interesting observation concerns the statistics of the Krylov angles. The Krylov angles deeper in the chain are found to have a narrower distribution, well approximated by a Gaussian. These results are discussed in Appendix \ref{App: Angle statistics}. An interesting observation is that the width of the distribution decreases slowly $\sim j^{-0.2}$, strongly suggesting a slow logarithmic narrowing with recursion step $j$. This feature  will be explored in more detail in future studies.

\begin{figure}
    \centering
    \includegraphics[width=0.4\textwidth]{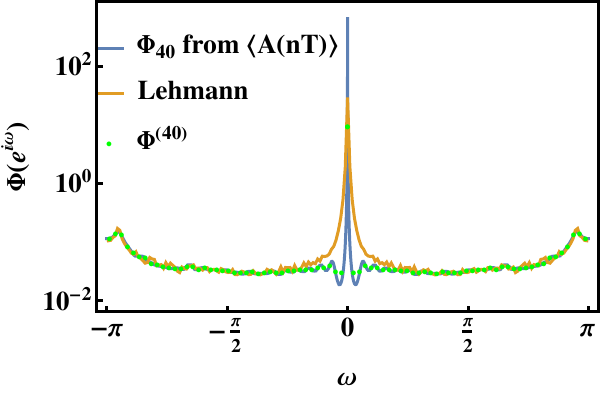}
    \caption{Comparison of different methods for extracting the disorder-averaged spectral function. The bottom two panels of Fig.~\ref{Fig: Anderson-5}, corresponding to the  Bernstein--Szeg\H{o} approximation with $k_* = 40$ ($\Phi_{40}$) and the truncated Fourier transform with $n_* = 40$ ($\Phi^{(40)}$), are plotted together with the results obtained from the Lehmann representation \eqref{Eq: Lehmann}, the latter averaged over $5000$ realizations. All three methods show good agreement, while the Bernstein--Szeg\H{o} approximation more efficiently captures the delta-function peak at $\omega = 0$.
    }
    \label{Fig: Spectral comparison}
\end{figure}

\begin{figure*}
    \centering
    \includegraphics[width=0.4\textwidth]{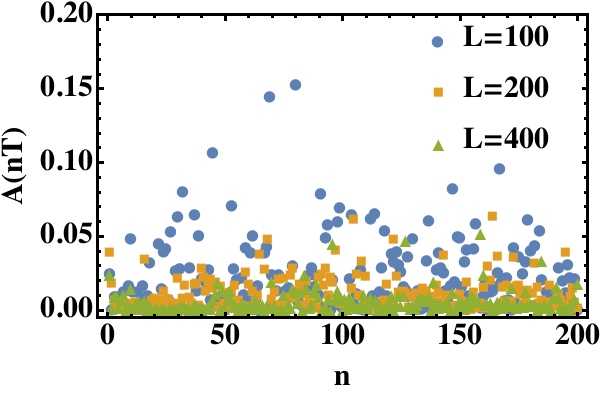}
    \includegraphics[width=0.4\textwidth]{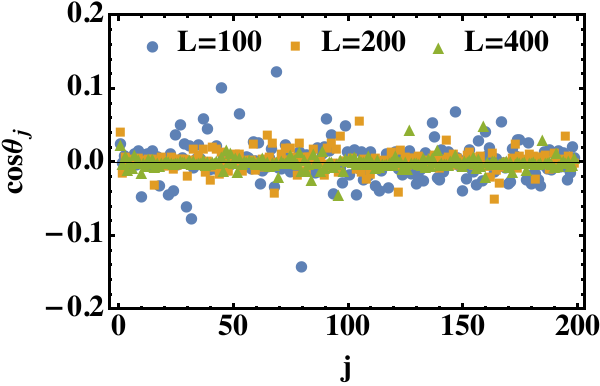}
    \includegraphics[width=0.4\textwidth]{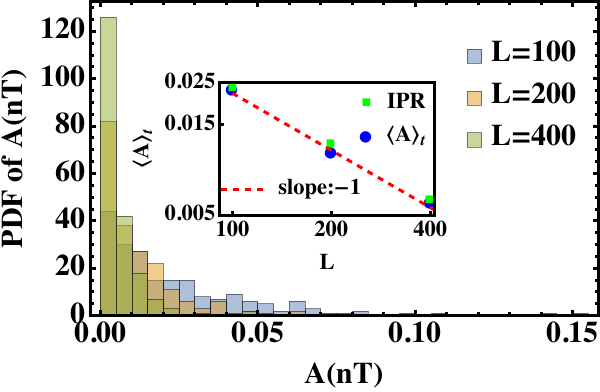}
    \includegraphics[width=0.4\textwidth]{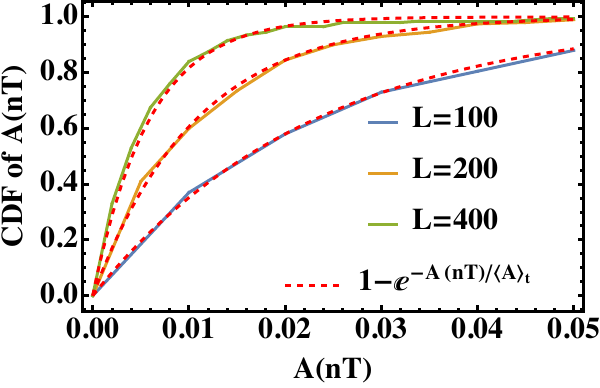}

    \caption{Delocalized phase of the Aubry--André model with $h = 1$, $T = 10$, and system sizes $L = 100, 200, 400$. Top panels: autocorrelation (left), and the corresponding Krylov angles (right). Bottom panels: PDF of the autocorrelation (left) and the corresponding CDF (right). Both the PDF and CDF are computed from the first 200 time steps of the autocorrelation. }
    \label{Fig: AA delocalized}
\end{figure*}

\section{Aubry–Andr\'e Model}\label{AAmodel}

For the Aubry--André (AA) model~\cite{aubry1980analyticity,dominguez2019aubry}, the local fields follow a quasi-periodic potential,
\begin{align}
    h_j =h\cos{(2\pi\beta(j-1))},\ \beta = \frac{1+\sqrt{5}}{2}.
\end{align}
In contrast to the 1D Anderson model, the AA model exhibits a localization--delocalization transition in one dimension: the system is localized for $h > 2$ and delocalized for $h < 2$. We present results in the localized phase, delocalized phase, and at the critical point. Moreover, we examine the autocorrelation more carefully in the delocalized phase and at the critical point, where it decays but continues to exhibit fluctuations at late times due to finite-size effects.

According to \eqref{Eq: A return}, the autocorrelation considered in this work is equivalent to the return probability. In the delocalized phase and for sufficiently large stroboscopic time step $T$, the state $|\phi(nT)\rangle$ is expected to behave effectively as a random state. If we average the autocorrelation over a long time interval, the resulting mean value, denoted by $\langle A \rangle_t$, is expected to coincide with the inverse participation ratio (IPR),
\begin{align}
    \langle A \rangle_t  
    &= \lim_{n_* \to \infty} \frac{1}{n_*} \sum_{n = 1}^{n_*} \left| \langle \phi |\phi(nT) \rangle \right|^2 \nonumber \\
    &= \lim_{n_* \to \infty} \frac{1}{n_*} \sum_{n = 1}^{n_*} \sum_{\epsilon_k,\epsilon_l} 
    |\langle \phi | \epsilon_k \rangle|^2 \, |\langle \phi | \epsilon_l \rangle|^2 
    e^{-i n (\epsilon_k - \epsilon_l)} \nonumber \\
    &= \sum_{\epsilon_k} |\langle \phi | \epsilon_k \rangle|^4 \propto L^{-D},
\end{align}
where $|\epsilon_k\rangle$ are eigenstates of $U$, defined by 
$U|\epsilon_k\rangle = e^{-i\epsilon_k}|\epsilon_k\rangle$, and $D$ is the scaling dimension.

\begin{figure*}
    \centering
    \includegraphics[width=0.4\textwidth]{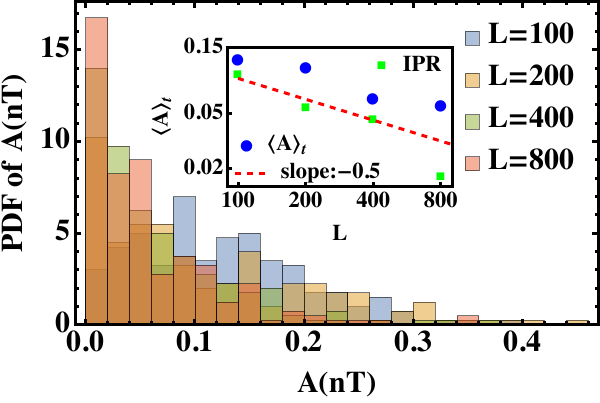}
    \includegraphics[width=0.4\textwidth]{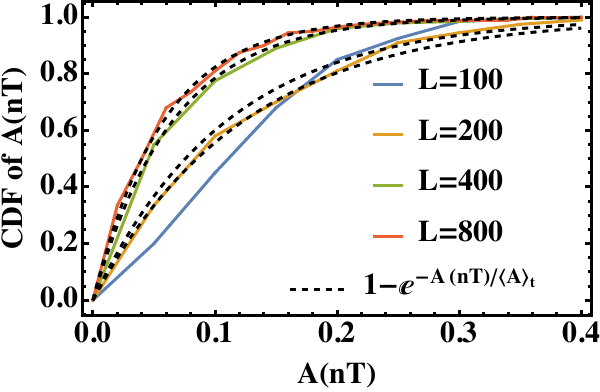}
    \includegraphics[width=0.4\textwidth]{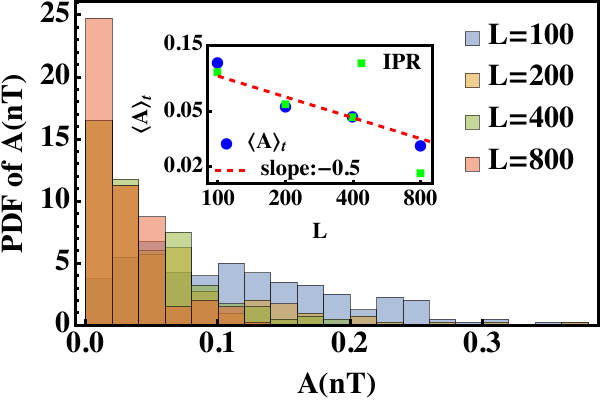}
    \includegraphics[width=0.4\textwidth]{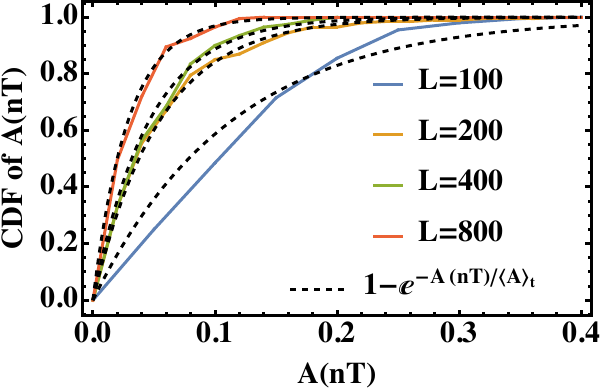}

    \caption{The Aubry--André model at the critical point ($h = 2$) for system sizes $L = 100, 200, 400, 800$ with $T = 10$ (top panels) and $T = 100$ (lower panels). In contrast to Fig.~\ref{Fig: AA delocalized}, $\langle A \rangle_t$ requires a larger stroboscopic time step $T$ to approach the IPR result (compare top and bottom insets of left panels). Nevertheless, the CDF already shows good agreement with the Porter--Thomas distribution for larger system sizes. The scaling of $\langle A \rangle_t$ is qualitatively consistent with the multifractal scaling at the transition point, $D \approx 0.5$ (left panel insets).}
    \label{Fig: AA crital}
\end{figure*}

For many-body chaotic systems, the late-time state is expected to behave as a Haar-random state and to follow a Porter--Thomas distribution~\cite{porter1956fluctuations,mullane2020sampling,claeys2025fock}. Although the AA model is non-interacting and its unitary dynamics is not exactly Haar random, the state in the delocalized phase exhibits behavior similar to that of random states in the late-time dephasing regime. Therefore, within this late-time dephasing window, we expect that the statistics of $A(nT)$ approximately follows a Porter--Thomas distribution. We will show this to indeed be the case. 
In particular, the probability density function (PDF) of $A(nT)$ within such a time interval exhibits an exponential distribution with mean $\langle A \rangle_t$. In practice, it is numerically more convenient to fit the cumulative distribution function (CDF), as the PDF is sensitive to histogram binning when the dataset is limited. The CDF of the autocorrelation is given by
\begin{align}
   \text{CDF of } A(nT): 1 - e^{-\frac{A(nT)}{\langle A \rangle_t}},
\end{align}
where $\langle A \rangle_t$ is obtained from the numerical time average and no additional fitting parameters are required.

We begin by considering the system in the delocalized phase ($h < 2$). In Fig.~\ref{Fig: AA delocalized}, we present numerical results for the autocorrelation (upper left panel) over the first 200 time steps, using a stroboscopic time step $T = 10$ with $h = 1$, for system sizes $L = 100, 200, 400$. The corresponding Krylov angles are shown in the upper right panel. The PDF of the autocorrelation (lower left panel) shows that the mean values $\langle A \rangle_t$ are consistent with the IPR, indicating a scaling dimension $D = 1$. Since the system is non-interacting and delocalized, $\langle A \rangle_t$ is expected to scale inversely with the system size. The corresponding CDF (lower right panel) exhibits good agreement with the Porter--Thomas distribution. The details of approximate spectral functions are presented in the Appendix \ref{App: AA numerical details}.

At the critical point ($h = 2$), the numerical analysis becomes more subtle for relatively small system sizes. In Fig.~\ref{Fig: AA crital}, we present the PDF and CDF of autocorrelations at the transition for system sizes $L = 100, 200, 400, 800$ and for two stroboscopic time steps, $T = 10$ (first row) and $T = 100$ (second row). The corresponding autocorrelation data are presented in Appendix~\ref{App: AA numerical details}. In contrast to Fig.~\ref{Fig: AA delocalized}, $T = 10$ is not sufficiently large to probe the late-time dynamics, and $\langle A \rangle_t$ deviates from the IPR result (see the upper left panel of Fig.~\ref{Fig: AA crital}). Nevertheless, the Porter--Thomas distribution provides a good description for larger system sizes. In the lower panels of Fig.~\ref{Fig: AA crital}, we use a larger stroboscopic time step, $T = 100$. In this case, $\langle A \rangle_t$ is closer to the IPR prediction, with only the $L = 100$ data still showing noticeable deviations from the Porter--Thomas distribution.

It is well known that the AA model exhibits multifractality at the transition point \cite{hiramoto1989scaling,wu2021fractal,wu2022aubry}, with numerical studies suggesting a scaling dimension $D \approx 0.5$ \cite{tang1986global,piechon1996anomalous}. However, accurately extracting this scaling requires more careful numerical analysis and sufficiently large system sizes. For instance, a noticeable drop in the IPR is observed for $L = 800$. In the insets of the left panels of Fig.~\ref{Fig: AA crital}, we include a red dashed line corresponding to $D = 0.5$. The data for $\langle A \rangle_t$ approximately follow this scaling, providing evidence for multifractal behavior at the transition.

\begin{figure*}
    \centering
    \includegraphics[width=0.3\textwidth]{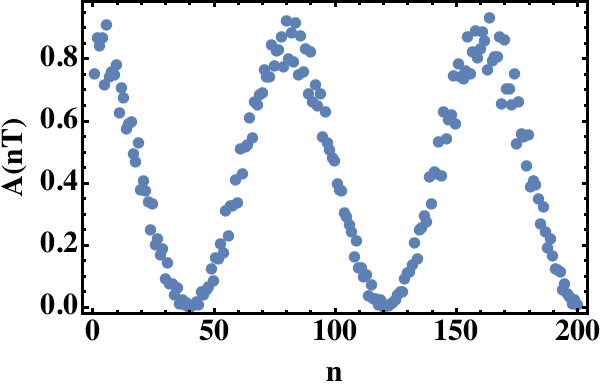}
    \includegraphics[width=0.32\textwidth]{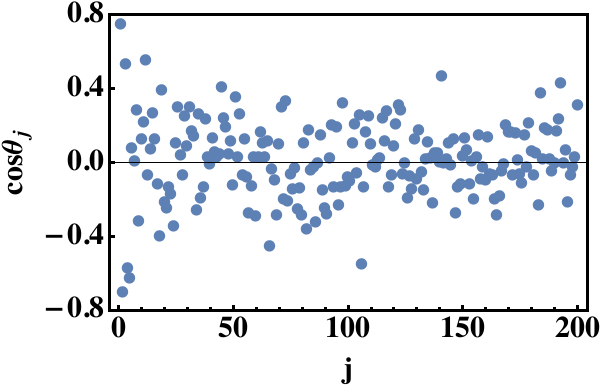}
    \includegraphics[width=0.3\textwidth]{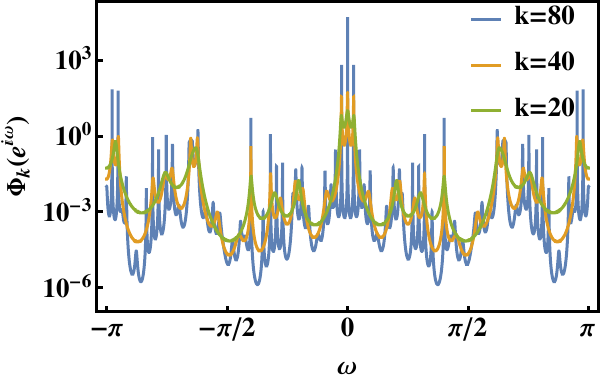}
    \caption{Localized phase of the Aubry--André model with $h = 3$, $T = 10$, and system size $L = 200$. Left panel: the autocorrelation exhibits persistent periodic oscillations. Middle panel: The corresponding Krylov angles. Right panel: approximate spectral function obtained from the Bernstein--Szeg\H{o}
 approximation, where pronounced peaks indicate long-lived modes in this phase.}
    \label{Fig: AA localized}
\end{figure*}

\begin{figure*}
    \centering
    \includegraphics[width=0.3\textwidth]{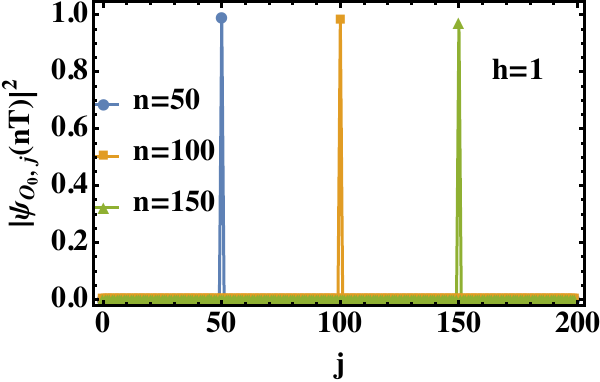}
    \includegraphics[width=0.3\textwidth]{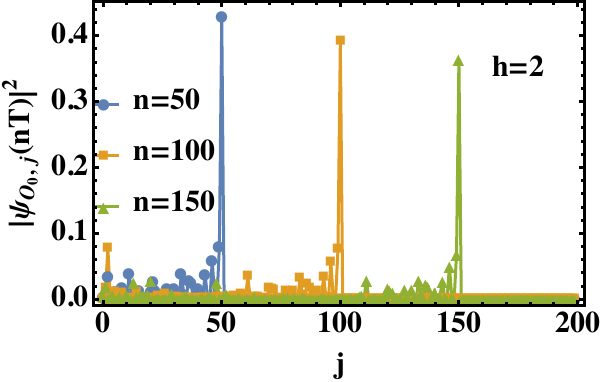}
    \includegraphics[width=0.3\textwidth]{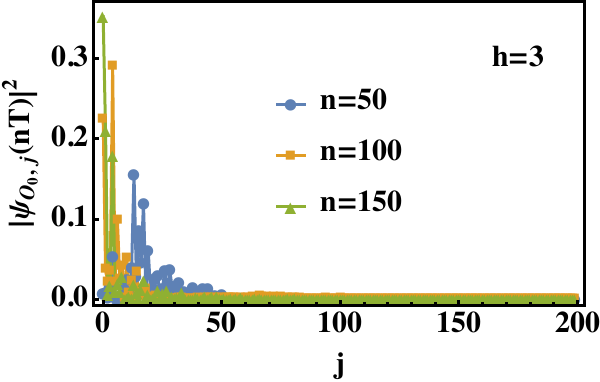}
    
    \caption{The modulus squared of the wave function in Krylov space at different time steps $n = 50, 100, 150$ with $L=200,\ T=10$. From left to right, three disorder strengths are considered: $h = 1$ (delocalized), $h = 2$ (critical point), and $h = 3$ (localized).}
    \label{Fig: AA spreading}
\end{figure*}

In the localized phase ($h > 2$), the autocorrelation does not decay. In Fig.~\ref{Fig: AA localized}, we present results for $h = 3$ and $L = 200$. The autocorrelation (left panel) remains finite at long times and exhibits persistent oscillations. The corresponding Krylov angles, shown in the middle panel, are computed using the Arnoldi iteration, as $\cos\theta_j$ exhibits strong fluctuations. The approximate spectral function displays pronounced peaks with large amplitudes, indicating the presence of long-lived modes in the dynamics. In particular, the peak located at $\omega = 0$ reflects the long-time oscillations observed in the autocorrelation.

Finally, we probe the phase transition of the AA model through the dynamics in Krylov space. In this framework, the dynamics can be viewed as the spreading of the initial operator $O_0$ along the Krylov chain, where each site corresponds to an orthonormal operator. This perspective allows us to examine how the spreading of $O_0$ depends on the disorder strength $h$.

In Fig.~\ref{Fig: AA spreading}, we explicitly show the operator spreading in the Krylov chain for the delocalized phase ($h = 1$, left panel), the critical point ($h = 2$, middle panel), and the localized phase ($h = 3$, right panel). The spreading is characterized by the squared amplitude of the operator wave function as 
\begin{align}
    |\psi_{O_0,j}(nT)|^2 = |(\mathcal{O}_j | O_0(nT))|^2.
\end{align}
In the delocalized phase, the wave function is peaked around $j \sim n$, indicating ballistic propagation along the Krylov chain. At the critical point, the wavefunction spreads into the bulk but develops a tail. In the localized phase, the wavefunction remains confined near $j \sim 0$.

\begin{figure}[h!]
    \centering
    \includegraphics[width=0.4\textwidth]{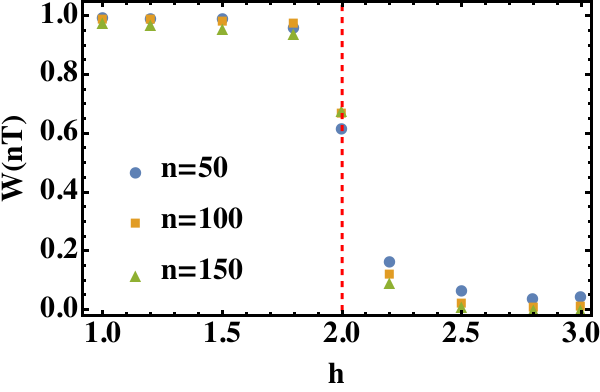}

    \caption{The wavefront weight $W(nT)$ defined in \eqref{Eq: wavefront} is computed for a range of disorder strengths $h \in [1,3]$ and for $L=200,\ T=10$. This quantity can probe the localization--delocalization transition, since the wave function is concentrated near the propagating front (initial sites) of the Krylov chain in the delocalized (localized) phase. The quantity $W$ exhibits a clear drop around $h \approx 2$.}
    \label{Fig: AA wave front}
\end{figure}

Motivated by this behavior, the magnitude of the propagating wavefront can serve as an indicator of the phase transition. To quantify this, we define the weight of the wavefront as the sum over the rightmost $1/5$ portion of the wave function,
\begin{align}
    W(nT) = \sum_{j = 4n/5}^{n} |(\mathcal{O}_j |O_0(nT))|^2.
    \label{Eq: wavefront}
\end{align}
As the system transitions from the delocalized to the localized phase, $W$ is expected to decrease from values close to $1$ to values near $0$. In Fig.~\ref{Fig: AA wave front}, we evaluate $W$ for several values of $h \in [1,3]$, observing a clear drop in the wavefront magnitude around $h = 2$, consistent with the location of the phase transition.

\section{Conclusions} \label{Concl}

Operator Krylov methods are emerging as  powerful methods for studying dynamics far out of equilibrium. We have shown that the Krylov methods for continuous time evolution have certain pitfalls that can be bypassed by
studying the dynamics stroboscopically, at discrete times, i.e, studying Floquet dynamics generated by the unitary $U=\exp(-i H T)$, where $H$ is the Hamiltonian of interest, and $T$ is an artificial time-period. We have applied this approach to the classic problem of Anderson localization and the single-particle localization-delocalization transition, both in 1D.

Despite being a single-particle problem, a Porter-Thomas distribution is found to capture the dynamics very well in the delocalized and the critical regime of the AA model. In addition, the critical point is captured well by operator spreading in Krylov space, with the emergence of a fractal scaling of quantities that depend on the IPR. The recursion procedure is found to naturally perform a renormalization group, with the distribution of Krylov parameters becoming narrower with recursion steps, approaching a "fixed point" where deep in the Krylov chain, with Krylov angles approaching $\pi/2$. 

There are several open questions. One is to generalize our study to higher spatial dimensions where the Anderson localization problem has more subtleties. 
The second is to understand better the fractal scaling at the critical point. The third is to understand why even a single particle problem should yield a Porter-Thomas distribution at long times even though the dynamics is not expected to be Haar-random to the same degree as a chaotic model. The fourth is to understand more quantitatively the statistics of the Krylov angles. The narrowing of their distribution is very slow, i.e., it fits a power law with a small coefficient. Understanding this feature better is an important direction, too. Finally, including interactions, is an important goal, and deferred to later studies. 

Quite generally with disorder, what are the right quantities to average in the context of Krylov space mappings is not fully understood. In particular, one needs to compare further the two ways of obtaining the Krylov parameters, one is from each disorder realization and then averaging it, and the second is to obtain them from already disorder averaged physical quantities such as the disorder-averaged autocorrelation function. The latter, in our study yielded the more physically sensible spectral functions, however the former  is natural to do when computational resources are limited, especially when performing traditional Hamiltonian Krylov computations. Understanding what physical properties the disorder averaged Krylov parameters might capture, making the first approach more sensible is an open question.

\emph{Acknowledgments}: The authors thank Sarang Gopalakrishnan, Clayton Peacock and Dries Sels for helpful discussions. This work was supported by the US Department of Energy, Office of
Science, Basic Energy Sciences, under Award No.~DE-SC0010821 (AM). HCY acknowledges support from the Max Planck Society.

\appendix

\section{Bilinear complex fermion representation}
\label{App: A}
In this section, we present relations for operators in the bilinear complex fermionic basis. We begin by deriving several identities involving fermionic creation and annihilation operators. Consider first the trace of $c_k^\dagger c_j$,
\begin{align}
    \text{Tr}[c_k^\dagger c_j] &= \text{Tr}[\delta_{jk} -  c_j c_k^\dagger] = \delta_{jk}\text{Tr}[\mathbb{I}] - \text{Tr}[c_j c_k^\dagger],
\end{align}
where we have used the anticommutation relation $\{c_k^\dagger, c_j\} = \delta_{jk}$. By cyclicity of the trace, it follows that
\begin{align}
    \text{Tr}[c_k^\dagger c_j] = \frac{\delta_{jk}}{2}\text{Tr}[\mathbb{I}].
    \label{Eq: trace cc}
\end{align}
Next, we evaluate the trace of a four-fermion operator $c_k^\dagger c_j c_l^\dagger c_m$. We first apply the anticommutation relation to the last two operators,
\begin{align}
    \text{Tr}[c_k^\dagger c_j c_l^\dagger c_m] &= \text{Tr}[c_k^\dagger c_j (\delta_{ml} -c_m c_l^\dagger)]\nonumber\\
    &= \delta_{ml}\text{Tr}[c_k^\dagger c_j ] - \text{Tr}[c_k^\dagger c_j c_m c_l^\dagger].
\end{align}
Using \eqref{Eq: trace cc} in the first term and the anticommutation relation $\{c_j, c_m\}=0$ in the second term, we obtain
\begin{align}
    \text{Tr}[c_k^\dagger c_j c_l^\dagger c_m]  =\frac{\delta_{ml}\delta_{jk}}{2}\text{Tr}[\mathbb{I}] + \text{Tr}[c_k^\dagger c_m c_j c_l^\dagger].
\end{align}
We then apply the anticommutation relation once more to the first two operators in the remaining term,
\begin{align}
    &\text{Tr}[c_k^\dagger c_j c_l^\dagger c_m]\nonumber\\
    &= \frac{\delta_{ml}\delta_{jk}}{2}\text{Tr}[\mathbb{I}] + \text{Tr}[(\delta_{mk} - c_m c_k^\dagger) c_j c_l^\dagger] \nonumber\\
    &=\frac{\delta_{ml}\delta_{jk}}{2}\text{Tr}[\mathbb{I}] + \delta_{mk}\text{Tr}[c_j c_l^\dagger] - \text{Tr}[c_m c_k^\dagger c_j c_l^\dagger].
\end{align}
Finally, using \eqref{Eq: trace cc} and cyclicity of the trace in the last term, we obtain
\begin{align}
    \text{Tr}[c_k^\dagger c_j c_l^\dagger c_m] = \frac{1}{4}(\delta_{ml}\delta_{jk} + \delta_{mk}\delta_{jl})\text{Tr}[\mathbb{I}].
    \label{Eq: trace cccc}
\end{align}

We are now in a position to rewrite the inner product~\eqref{Eq: inner prod} in the bilinear complex fermionic basis. For two bilinear operators $A$ and $B$ of the form
\begin{align}
    A =\sum_{j,k}\Tilde{A}_{jk}c_j^\dagger c_k,\ B =\sum_{j,k}\Tilde{B}_{jk}c_j^\dagger c_k,
\end{align}
the first term in \eqref{Eq: inner prod} becomes
\begin{align}
    \frac{1}{\text{Tr}[\mathbb{I}]} \text{Tr}[A^\dagger B] &= \frac{1}{\text{Tr}[\mathbb{I}]}\sum_{j,k,l,m} \overline{\Tilde{A}_{jk}}\Tilde{B}_{lm} \text{Tr}[c_k^\dagger c_j c_l^\dagger c_m]\nonumber\\
    &= \frac{1}{4}\sum_{j,k,l,m} \overline{\Tilde{A}_{jk}}\Tilde{B}_{lm} \left(\delta_{mk}\delta_{jl} + \delta_{ml}\delta_{jk} \right)\nonumber\\
    &=\frac{1}{4} \left( \text{tr}[\Tilde{A}^\dagger \Tilde{B}] + \text{tr}[\Tilde{A}^\dagger] \text{tr}[\Tilde{B}] \right),
\end{align}
where \eqref{Eq: trace cccc} has been used in the second line, and $\text{tr}[\cdot]$ denotes the trace in the bilinear complex fermionic basis.

For the second term in \eqref{Eq: inner prod}, we find
\begin{align}
    &\frac{1}{\text{Tr}[\mathbb{I}]}\text{Tr}[A^\dagger] \frac{1}{\text{Tr}[\mathbb{I}]}\text{Tr}[B]\nonumber\\
    &= \frac{1}{(\text{Tr}[\mathbb{I}])^2} \left(\sum_{j,k}\overline{\Tilde{A}_{jk}}\text{Tr}[c_k^\dagger c_j ]\right)\left(\sum_{l,m}\Tilde{B}_{lm}\text{Tr}[c_l^\dagger c_m]\right)\nonumber\\
    &= \frac{1}{4} \left(\sum_{j,k}\overline{\Tilde{A}_{jk}}\delta_{jk} \right)\left(\sum_{l,m}\Tilde{B}_{lm}\delta_{lm}\right)\nonumber\\
    &= \frac{1}{4} \text{tr}[\Tilde{A}^\dagger] \text{tr}[\Tilde{B}],
\end{align}
where \eqref{Eq: trace cc} has again been used. Combining both contributions, we obtain
\begin{align}
    (A|B) &= \frac{1}{\text{Tr}[\mathbb{I}]}\text{Tr}[A^\dagger B] - \frac{1}{\text{Tr}[\mathbb{I}]^2}\text{Tr}[A^\dagger]\text{Tr}[B]\nonumber\\
    &=\frac{1}{4} \text{tr}[\Tilde{A}^\dagger\Tilde{B}].
    \label{Eq: inner prod bilinear}
\end{align}
 
Next, we derive the time evolution of operators in the bilinear complex fermionic basis. The commutator between the Hamiltonian and an operator is given by
\begin{align}
    \mathcal{L}_{H} O &\equiv [H,O] = \sum_{j,k,l,m} \Tilde{H}_{jk} \Tilde{O}_{lm} [c_j^\dagger c_k, c_l^\dagger c_m].
\end{align}
Expanding the commutator explicitly yields
\begin{align}
    [c_j^\dagger c_k, c_l^\dagger c_m] &= c_j^\dagger c_k c_l^\dagger c_m - c_l^\dagger c_m c_j^\dagger c_k \nonumber\\
    &= \delta_{kl} c_j^\dagger c_m - c_j^\dagger c_l^\dagger c_k c_m - \delta_{mj}c_l^\dagger c_k + c_l^\dagger c_j^\dagger c_m c_k \nonumber\\
    &=\delta_{kl} c_j^\dagger c_m - \delta_{mj}c_l^\dagger c_k,
\end{align}
where all quartic terms cancel. Hence,
\begin{align}
    \mathcal{L}_{H} O 
    &= \sum_{j,k,l,m} \Tilde{H}_{jk} \Tilde{O}_{lm} (\delta_{kl}c_j^\dagger c_m - \delta_{jm}c_l^\dagger c_k)\nonumber\\
    &= \sum_{j,k,m} \Tilde{H}_{jk}\Tilde{O}_{km} c_j^\dagger c_m - \sum_{j,k,l} \Tilde{H}_{jk} \Tilde{O}_{lj}c_l^\dagger c_k\nonumber\\
    &=\sum_{j,k,m} (\Tilde{H}_{jk} \Tilde{O}_{km} - \Tilde{O}_{jk} \Tilde{H}_{km}) c_j^\dagger c_m\nonumber\\
    &\equiv  \sum_{j,m}(\mathcal{L}_{\Tilde{H}}\Tilde{O})_{jm} c_j^\dagger c_m,
\end{align}
where $\mathcal{L}_{\Tilde{H}}\Tilde{O}$ denotes the matrix commutator in the bilinear fermionic basis.

Therefore, the time evolution of an operator is
\begin{align}
    O(n) &= e^{iHnT} O e^{-iHnT} \nonumber\\
    &= \sum_l \frac{(inT\mathcal{L}_H)^l O}{l!}\nonumber\\
    &= \sum_l \sum_{j,k}\left(\frac{(inT\mathcal{L}_{\Tilde{H}})^l \Tilde{O}}{l!}\right)_{jk} c_j^\dagger c_k\nonumber\\
    &=
    \sum_{j,k} \left( e^{i\Tilde{H}n T} \Tilde{O} e^{-i\Tilde{H}n T} \right)_{jk}c_j^\dagger c_k.
    \label{Eq: O(n) bilinear}
\end{align}
Combining~\eqref{Eq: inner prod bilinear} and~\eqref{Eq: O(n) bilinear}, we recover the autocorrelation function~\eqref{Eq: auto bilinear}.

\begin{widetext}
\section{Stroboscopic time-step dependence and comparison with the Arnoldi iteration for the Anderson model}
\label{App: Anderson numerical details}

We further support the results in Fig.~\ref{Fig: Anderson-1} by providing additional details on the dependence on the stroboscopic time step and the use of the Arnoldi iteration for a strong-disorder example. In Fig.~\ref{Fig: Anderson-3}, we compare numerical results for stroboscopic time steps $T = 1.5$ and $T = 30$. For larger $T$, the early-time transients in the autocorrelation are smoothed out, while the Krylov angles exhibit the same asymptotic power-law behavior.

In Fig.~\ref{Fig: Anderson-2}, we present additional data for $h = 10$ obtained using the Arnoldi iteration (see Sec.~\ref{Sec: Gram-Schmidt}). These results confirm that the asymptotic behavior, $\langle \cos\theta_{2j-1} \rangle \sim 1/\sqrt{j}$, in the strong-disorder regime is already well captured by the limited data shown in Fig.~\ref{Fig: Anderson-1}.
\begin{figure}[h!]
    \centering
    \includegraphics[width=0.4\textwidth]{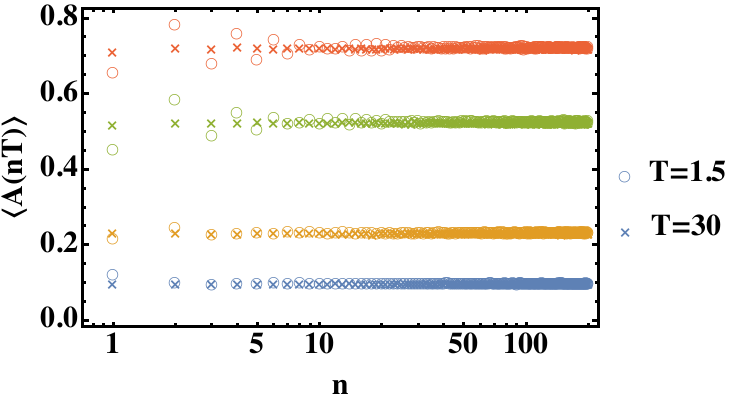}
    \includegraphics[width=0.4\textwidth]{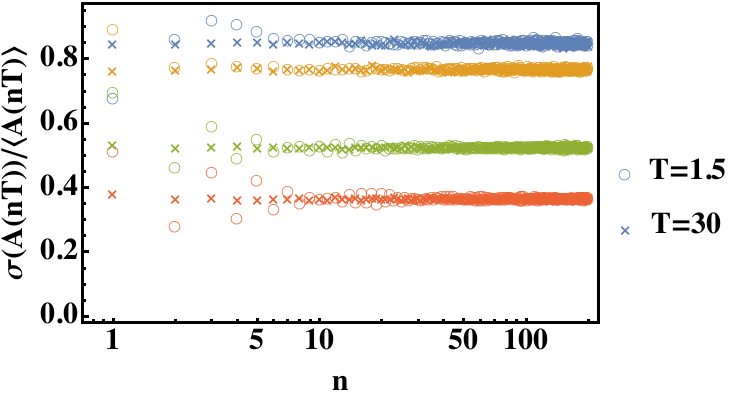}
    
    \includegraphics[width=0.4\textwidth]{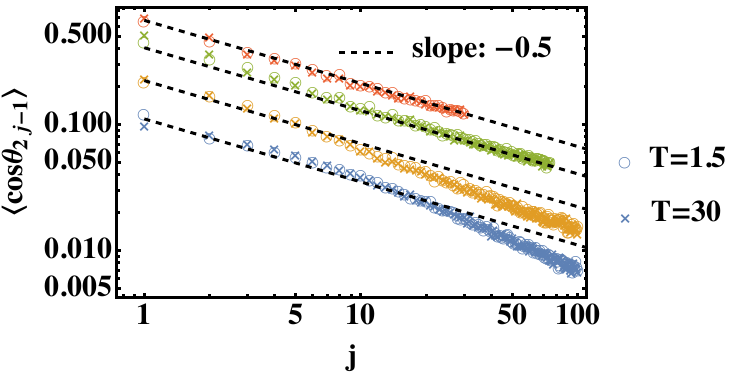}
    \includegraphics[width=0.4\textwidth]{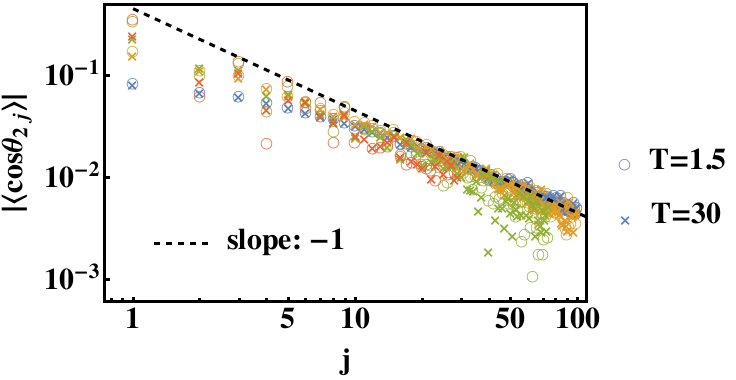}

    \caption{Comparison between $T = 1.5$ (same as Fig.~\ref{Fig: Anderson-1}) and $T = 30$ for the Anderson model. Top panels, the $x$-axes are on a log-scale. Bottom panels are on a log-log scale. 
    At late times, the numerical results are insensitive to $T$.}
    \label{Fig: Anderson-3}
\end{figure}

\begin{figure}[h!]
    \centering
    \includegraphics[width=0.4\textwidth]{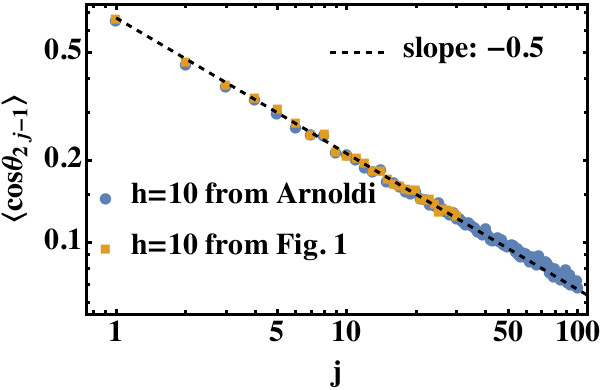}
    \includegraphics[width=0.4\textwidth]{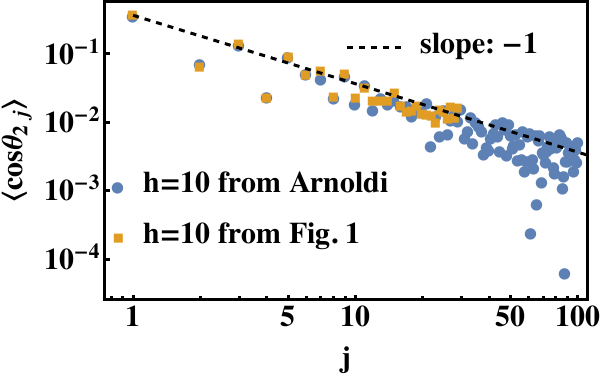}

    \caption{Comparison of numerical results for the disorder-averaged cosine of the Krylov angles obtained from the Arnoldi iteration and from the moment method, the latter shown in Fig.~\ref{Fig: Anderson-1}, for $h = 10$, $T = 1.5$, and $15{,}000$ realizations. The data are presented on a log--log scale, with $\langle \cos\theta_{2j-1} \rangle$ and $\langle \cos\theta_{2j} \rangle$ shown in the left and right panels respectively. The moment method numerics become less stable at strong disorder, since the angles do not approach $\pi/2$ sufficiently rapidly, causing the denominator in \eqref{Eq: moment algo} to become small.}
    \label{Fig: Anderson-2}
\end{figure}

\section{Dips in the spectral functions obtained from the disorder averaged Krylov angles}
\label{App: spectral dips}
In this appendix, we take a closer look at the phenomena in Fig.~\ref{Fig: Anderson-5} top right panel, where there are deep dips in the spectral function that flank the delta-function peak at zero frequency, with similar dips appearing concurrently at the zone boundaries $\omega=\pm \pi$. This only occurs when the spectral function is computed from the disorder-averaged Krylov angles, and is particularly enhanced at strong disorder ($h=10$ in Fig.~\ref{Fig: Anderson-5}). 
Since the approximate spectral function is related to the OPUC through \eqref{Eq: Bernstein approx}, we focus on the behavior of $P_k(e^{i\omega})$ in the following discussion.

\begin{figure}[h!]
    \centering
    \includegraphics[width=0.4\textwidth]{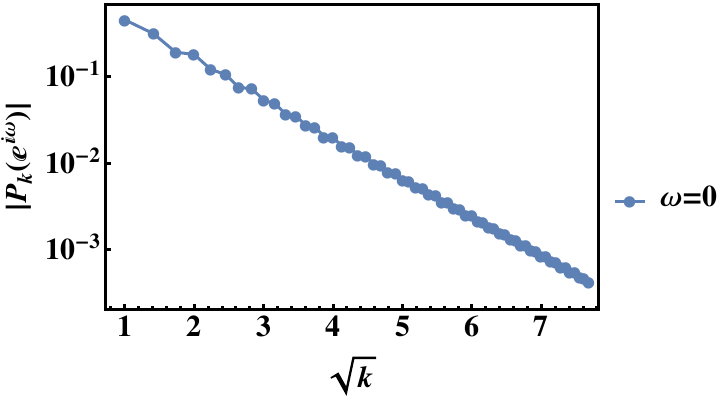}
    \includegraphics[width=0.4\textwidth]{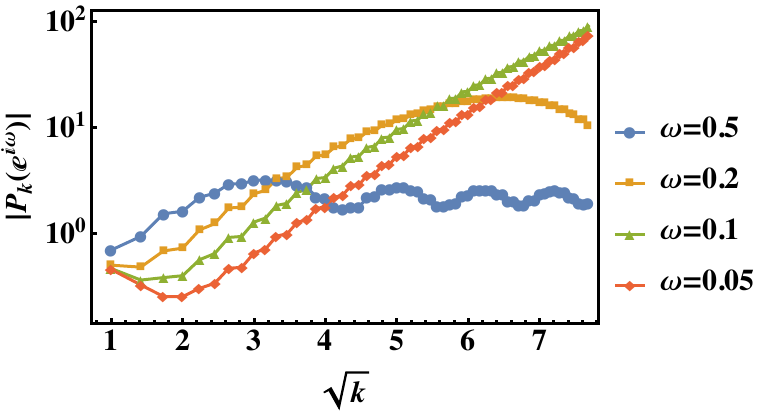}

    \caption{Numerical results for $|P_k(e^{i\omega})|$ at $\omega = 0, 0.05, 0.1, 0.2, 0.5$, obtained from disorder-averaged Krylov angles for the Anderson model with $h=10$. At $\omega = 0$ (left panel), $\ln |P_k(e^{i\omega})| \propto -\sqrt{k}$, consistent with the divergence of the spectral function at $\omega = 0$. Near $\omega = 0$ (right panel), $|P_k(e^{i\omega})|$ exhibits oscillatory convergence for $k > k_c(\omega)$. Qualitatively, $k_c(\omega)$ increases as $\omega$ decreases. For sufficiently small $\omega$, $|P_k(e^{i\omega})|$ shows a  growth (after an initial transient) characterized by $\ln |P_k(e^{i\omega})| \propto \sqrt{k}$. It is this growth which causes a pronounced dip in the spectral function in the vicinity of the $\delta$-function peak at $\omega=0$.}
    \label{Fig: spectral dip}
\end{figure}

For the spectral function near $\omega = 0$, we explicitly present numerical results for $|P_k(e^{i\omega})|$ at $\omega = 0, 0.05, 0.1, 0.2, 0.5$ in Fig.~\ref{Fig: spectral dip}. At $\omega = 0$ (left panel), we observe that $\ln |P_k(1)| \propto -\sqrt{k}$, implying that $\Phi_k(1)$ diverges as $k \to \infty$ and therefore gives rise to a delta-function contribution at $\omega = 0$. 

For frequencies close to $\omega = 0$ (right panel), $|P_k(e^{i\omega})|$ exhibits oscillatory convergence for $\omega = 0.5$. As $\omega$ decreases, however, the onset of this oscillatory behavior shifts to larger values of $k$. In particular, for smaller frequencies such as $\omega = 0.2$, one first observes an early-growth regime characterized by $\ln |P_k(e^{i\omega})| \propto \sqrt{k}$, followed by a crossover to oscillatory convergence at some scale $k = k_c(\omega)$. This early-growth regime becomes even more pronounced for $\omega = 0.05$ and $0.1$. Numerical results in Fig.~\ref{Fig: spectral dip} indicate that $k_c(\omega)$ increases as $\omega$ decreases. Therefore, the true spectral function near $\omega = 0$ can be estimated by the approximate spectral function evaluated at $k_c(\omega)$,
\begin{align}
    \Phi(e^{i\omega}) \approx \frac{1}{2\pi}   \frac{1}{|P_{k_c(\omega)}(e^{i\omega})|^2}.\label{Eq: spectral estimate}
\end{align}
Combining the numerical behavior in the right panel of Fig.~\ref{Fig: spectral dip} with \eqref{Eq: spectral estimate} suggests a suppression of the spectral function, i.e., the formation of a dip near $\omega = 0$. Below, we derive the square-root dependence on $k$ observed in Fig.~\ref{Fig: spectral dip}.

We first consider the case of $\omega = 0$, for which the Szeg\H{o} recurrence relation~\eqref{Eq: Szego recurrence} simplifies to
\begin{align}
    \begin{pmatrix}
        P_{k}(1)\\
        P_{k}^*(1)
    \end{pmatrix}=  \frac{1}{\sin\theta_k}
    \begin{pmatrix}
        1 & (-1)^k \cos\theta_k \\
        (-1)^k \cos\theta_k &  1
    \end{pmatrix}
    \begin{pmatrix}
        P_{k-1}(1)\\
        P_{k-1}^*(1)
    \end{pmatrix}.
\end{align}
Since $\theta_k \in  [0,\pi]$, it is straightforward to show that
\begin{align}
    \frac{1}{\sin\theta_k}
    \begin{pmatrix}
        1 & (-1)^k \cos\theta_k \\
        (-1)^k \cos\theta_k &  1
    \end{pmatrix}
    \begin{pmatrix}
        1\\
        1
    \end{pmatrix} = \frac{1+(-1)^k\cos\theta_k}{\sin\theta_k}\begin{pmatrix}
        1\\
        1
    \end{pmatrix} =\sqrt{\frac{1+(-1)^k\cos\theta_k}{1-(-1)^k\cos\theta_k}}\begin{pmatrix}
        1\\
        1
    \end{pmatrix}.
\end{align}
Therefore,
\begin{align}
    \begin{pmatrix}
        P_{2k}(1)\\
        P_{2k}^*(1)
    \end{pmatrix}= \left(\prod_{j=1}^{k}\sqrt{\frac{1+\cos\theta_{2j}}{1-\cos\theta_{2j}}} \sqrt{\frac{1-\cos\theta_{2j-1}}{1+\cos\theta_{2j-1}}}\right) 
    \begin{pmatrix}
        1\\
        1
    \end{pmatrix} = \left|\frac{\psi_{2k+1}}{\psi_1}\right|\begin{pmatrix}
        1\\
        1
    \end{pmatrix},
    \label{Eq: OPUC at omega 0}
\end{align}
where \eqref{Eq: wave function} is used in the last step. According to \eqref{Eq: ave Krylov angles} and \eqref{Eq: wave function geo ave}, one obtains $\ln |P_k(1)| \propto -\sqrt{k}$, consistent with the left panel of Fig.~\ref{Fig: spectral dip}.

For $0 < \omega \ll 1$, the Szeg\H{o} recurrence relation~\eqref{Eq: Szego recurrence} can be approximated as 
\begin{align}
    \begin{pmatrix}
        P_{k}(e^{i\omega})\\
        P_{k}^*(e^{i\omega})
    \end{pmatrix} &\approx  \frac{1}{\sin\theta_k}\left[
    \begin{pmatrix}
        1 & (-1)^k \cos\theta_k \\
        (-1)^k \cos\theta_k &  1
    \end{pmatrix} +i\omega  \begin{pmatrix}
        1 & 0 \\
        (-1)^k \cos\theta_k &  0
        \end{pmatrix}\right]
    \begin{pmatrix}
        P_{k-1}(e^{i\omega})\\
        P_{k-1}^*(e^{i\omega})
    \end{pmatrix}\nonumber\\
    &\approx\frac{1}{\sin\theta_k}\left[
    \begin{pmatrix}
        1 & (-1)^k \cos\theta_k \\
        (-1)^k \cos\theta_k &  1
    \end{pmatrix} +i\omega  \begin{pmatrix}
        1 & 0 \\
        0 &  0
        \end{pmatrix}\right]
    \begin{pmatrix}
        P_{k-1}(e^{i\omega})\\
        P_{k-1}^*(e^{i\omega})
    \end{pmatrix},
\end{align}
where we used $e^{i\omega} \approx 1 + i\omega$. In the second line, we retain only the leading matrix element of the second matrix, i.e, by using $|\cos{\theta_k}| \ll 1$, we have dropped it.   For convenience, we introduce
\begin{align}
    M_k = \frac{1}{\sin\theta_k}
    \begin{pmatrix}
        1 & (-1)^k \cos\theta_k \\
        (-1)^k \cos\theta_k &  1
    \end{pmatrix},\ m_k = \frac{i\omega}{\sin\theta_k}\begin{pmatrix}
        1 & 0 \\
        0 &  0
        \end{pmatrix}.
\end{align}
Acting with $m_k$ on $(1,1)^\intercal$ gives
\begin{align}
    m_k \begin{pmatrix}
        1\\
        1
    \end{pmatrix}
    = \frac{i\omega}{\sin\theta_k} \begin{pmatrix}
        1\\
        0
    \end{pmatrix}
    = \frac{i\omega}{2\sin\theta_k}\left[
    \begin{pmatrix}
        1\\
        1
    \end{pmatrix}+
    \begin{pmatrix}
        1\\
        -1
    \end{pmatrix}
    \right].
\end{align}
Thus, the perturbation generates the additional vector $(1,-1)^\intercal$. Acting with $M_k$ on this vector yields
\begin{align}
    M_k \begin{pmatrix}
        1\\
        -1
    \end{pmatrix} = \frac{1-(-1)^k\cos\theta_k}{\sin\theta_k}\begin{pmatrix}
        1\\
        -1
    \end{pmatrix} =\sqrt{\frac{1-(-1)^k\cos\theta_k}{1+(-1)^k\cos\theta_k}}\begin{pmatrix}
        1\\
        -1
    \end{pmatrix}.
\end{align}
Hence, $(1,1)^\intercal$ and $(1,-1)^\intercal$ are eigenvectors of $M_k$, with eigenvalues whose product equals unity.

We now examine the leading-order behavior of the OPUC,
\begin{align}
    \begin{pmatrix}
        P_{2k}(e^{i\omega})\\
        P_{2k}^*(e^{i\omega})
    \end{pmatrix} &\approx \left(\prod_{j=1}^{2k}(M_j+m_j) \right)\begin{pmatrix}
        1\\
        1
    \end{pmatrix}\nonumber\\
    &= \left(\prod_{j=1}^{2k}M_j \right)\begin{pmatrix}
        1\\
        1
    \end{pmatrix} + \left[\sum_{l=1}^{2k}\left(\prod_{p=l+1}^{2k}M_p \right)m_l\left(\prod_{j=1}^{l-1}M_j \right) \right]\begin{pmatrix}
        1\\
        1
    \end{pmatrix}+ \ldots
    \label{Eq: OPUC expansion}
\end{align}
The first term is stretched-exponentially small according to \eqref{Eq: OPUC at omega 0}. The dominant contribution in the summation arises from
\begin{align} 
    \left(\prod_{j=2}^{2k}M_j \right)m_1\begin{pmatrix}
        1\\
        1
    \end{pmatrix} &= \left(\prod_{j=2}^{2k}M_j \right)\frac{i\omega}{2\sin\theta_1}\left[
    \begin{pmatrix}
        1\\
        1
    \end{pmatrix}+
    \begin{pmatrix}
        1\\
        -1
    \end{pmatrix}\right]\nonumber\\
    &=\left|\frac{\psi_{2k+1}}{\psi_1}\right|\sqrt{\frac{1+\cos\theta_1}{1-\cos\theta_1}}\frac{i\omega}{2\sin\theta_1}\begin{pmatrix}
        1\\
        1
    \end{pmatrix} +
    \left|\frac{\psi_1}{\psi_{2k+1}}\right|\sqrt{\frac{1-\cos\theta_1}{1+\cos\theta_1}}\frac{i\omega}{2\sin\theta_1}\begin{pmatrix}
        1\\
        -1
    \end{pmatrix},
\end{align}
where the second term dominates stretched-exponentially. The summation in \eqref{Eq: OPUC expansion} does not affect the leading stretched-exponential behavior.  Using \eqref{Eq: ave Krylov angles} and \eqref{Eq: wave function geo ave}, we therefore obtain
\begin{align}
    \ln |P_k(e^{i\omega})| \propto \sqrt{k},\ \forall\ k \leq k_c(\omega),
\end{align}
in agreement with the numerical results shown in the right panel of Fig.~\ref{Fig: spectral dip}. The growth terminates near $k = k_c(\omega)$ due to cancellations from higher-order terms of $m_k$ in \eqref{Eq: OPUC expansion}.

Finally, we consider $\omega = \pi$, for which the Szeg\H{o} recurrence relation becomes
\begin{align}
    \begin{pmatrix}
        P_{k}(-1)\\
        P_{k}^*(-1)
    \end{pmatrix}=  \frac{1}{\sin\theta_k}
    \begin{pmatrix}
        -1 & (-1)^k \cos\theta_k \\
        -(-1)^k \cos\theta_k &  1
    \end{pmatrix}
    \begin{pmatrix}
        P_{k-1}(-1)\\
        P_{k-1}^*(-1)
    \end{pmatrix}.
\end{align}
One can straightforwardly verify that
\begin{subequations}
\begin{align}
    &\frac{1}{\sin\theta_k}
    \begin{pmatrix}
        -1 & (-1)^k \cos\theta_k \\
        -(-1)^k \cos\theta_k &  1
    \end{pmatrix}
    \begin{pmatrix}
        1\\
        1
    \end{pmatrix} =-\sqrt{\frac{1-(-1)^k\cos\theta_k}{1+(-1)^k\cos\theta_k}}\begin{pmatrix}
        1\\
        -1
    \end{pmatrix},\\
    &\frac{1}{\sin\theta_k}
    \begin{pmatrix}
        -1 & (-1)^k \cos\theta_k \\
        -(-1)^k \cos\theta_k &  1
    \end{pmatrix}
    \begin{pmatrix}
        1\\
        -1
    \end{pmatrix} =-\sqrt{\frac{1+(-1)^k\cos\theta_k}{1-(-1)^k\cos\theta_k}}\begin{pmatrix}
        1\\
        1
    \end{pmatrix}.
\end{align}
\end{subequations}
Therefore, 
\begin{align}
    \begin{pmatrix}
        P_{2k}(-1)\\
        P_{2k}^*(-1)
    \end{pmatrix}= \left(\prod_{j=1}^{k}\sqrt{\frac{1+\cos\theta_{2j}}{1-\cos\theta_{2j}}} \sqrt{\frac{1+\cos\theta_{2j-1}}{1-\cos\theta_{2j-1}}}\right) 
    \begin{pmatrix}
        1\\
        1
    \end{pmatrix}\sim e^{\sum_{j=1}^{k} \left(  \cos\theta_{2j-1} + \cos\theta_{2j}\right)} \begin{pmatrix}
        1\\
        1
    \end{pmatrix},
    \label{Eq: OPUC at pi}
\end{align}
where the sign of the odd Krylov-angle contribution is opposite to that in \eqref{Eq: OPUC at omega 0}. According to \eqref{Eq: ave Krylov angles}, the asymptotics of $P_{2k}(-1)$ is dominated by $\cos\theta_{2j-1}$: $P_{2k}(-1) \sim e^{2\eta_o\sqrt{k}} \rightarrow \infty$. Consequently, the spectral function develops a dip at $\omega = \pi$.

In summary, the dip in the spectral function near $\omega = 0$ in Fig.~\ref{Fig: Anderson-2} originates from the early growth of the OPUC, which is directly tied to the localization of the $0$-mode wavefunction in Krylov space. This argument becomes more subtle for power-law localized wavefunctions, since the summation in \eqref{Eq: OPUC expansion} may modify the leading power-law exponent. Nevertheless, the numerical results for $|P_k(e^{i\omega})|$ shown in Fig.~\ref{Fig: spectral dip-2}, corresponding to the lower-left panel in Fig.~\ref{Fig: Anderson-5}, exhibit behavior similar to that in Fig.~\ref{Fig: spectral dip}. However, the oscillatory convergence of $|P_k(e^{i\omega})|$ occurs while it remains of order unity; namely, $k_c(\omega)$ is too small for a dip to develop in the spectral function in the vicinity of the delta-function peak. Moreover, as shown in Fig.~\ref{Fig: Anderson-4}, the odd and even contributions are equally important. Consequently, the cancellation of the Krylov angles in the exponent of \eqref{Eq: OPUC at pi} yields an OPUC of order unity as $k \rightarrow \infty$, and therefore no dip appears at $\omega = \pi$ in Fig.~\ref{Fig: Spectral comparison}. A cusp occurs near $\omega=\pi$ in Fig.~\ref{Fig: Spectral comparison} however, its variation is only about $0.1$. Moreover, the lower-left panel of Fig.~\ref{Fig: Anderson-5} shows that $|P_{2k}(-1)|$ saturates to a constant value at large $k$. 
This indicated that the behavior at $\omega=\pi$ is not governed by a leading asymptotic term, but depends on details. In particular,
the cancellation in \eqref{Eq: OPUC at pi} is delicate as $\cos\theta_{2j-1}\sim -\cos\theta_{2j}\sim 1/j$; a slight difference between the prefactors of the even and odd Krylov angles can cause $|P_{2k}(-1)|$ to either grow or decay as a power law.

\begin{figure}[h!]
    \centering
    \includegraphics[width=0.4\textwidth]{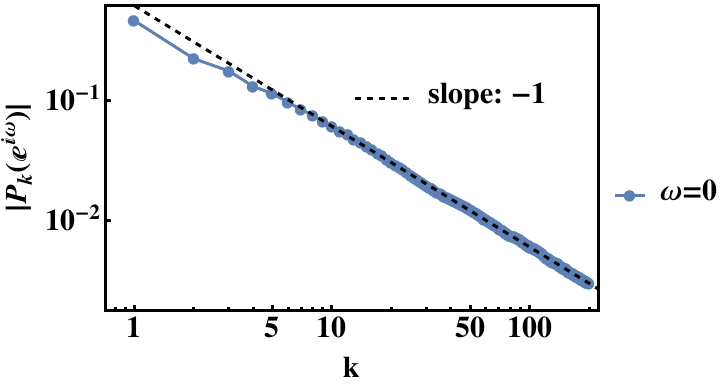}
    \includegraphics[width=0.4\textwidth]{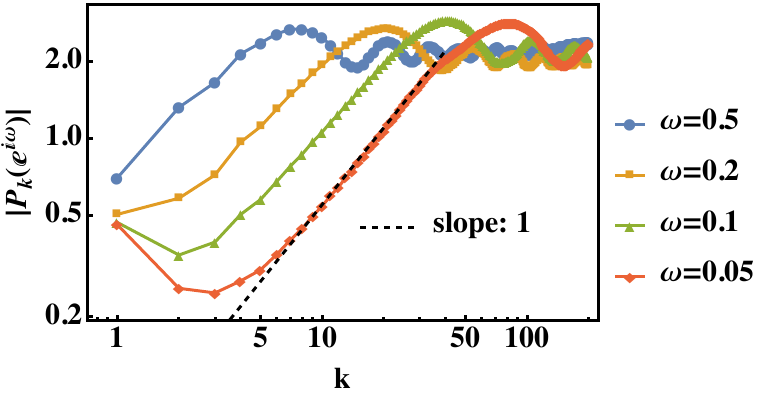}

    \caption{Numerical results for $|P_k(e^{i\omega})|$ at $\omega = 0, 0.05, 0.1, 0.2, 0.5$, obtained from the disorder-averaged autocorrelation function for the Anderson model with $h=10$. At $\omega = 0$ (left panel), $|P_k(e^{i\omega})| \propto 1/k$, consistent with the divergence of the spectral function at $\omega = 0$. Similar to Fig.~\ref{Fig: spectral dip}, $|P_k(e^{i\omega})|$ exhibits oscillatory convergence for $k > k_c(\omega)$ near $\omega = 0$ (right panel). For sufficiently small $\omega$, $|P_k(e^{i\omega})|$ initially grows as $|P_k(e^{i\omega})| \propto k$. However, this growth saturates at $|P_k(e^{i\omega})| \approx 2$, preventing the appearance of a dip in the spectral function.}
    \label{Fig: spectral dip-2}
\end{figure}

\section{Numerical results for the Lehmann representation of the spectral-function}
\label{App: Lehmann}
We present more detailed numerical results for the Lehmann representation of the spectral function in \eqref{Eq: Lehmann} for the Anderson model with $h = 10$. In Fig.~\ref{Fig: Lehmann}, we show results for two different width parameters, $\eta = \pi/L$ and $\eta = \pi/(2L)$ (see \eqref{Eq: wrapped Lorentzian}), in the left and right panels, respectively, for system sizes $L = 50, 100, 200$, averaged over $5000$ realizations. The peak at $\omega = 0$ becomes sharper for larger system sizes or smaller values of $\eta$, consistent with the emergence of a delta function in the thermodynamic limit. However, achieving a smooth spectrum at smaller $\eta$ requires averaging over a larger number of realizations to reduce statistical noise. 

\begin{figure}[h!]
    \centering
    \includegraphics[width=0.4\textwidth]{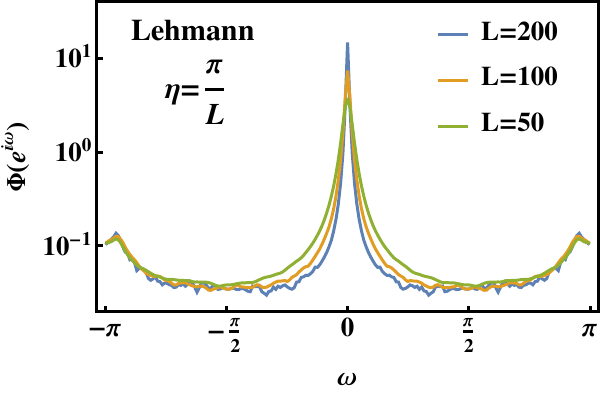}
    \includegraphics[width=0.4\textwidth]{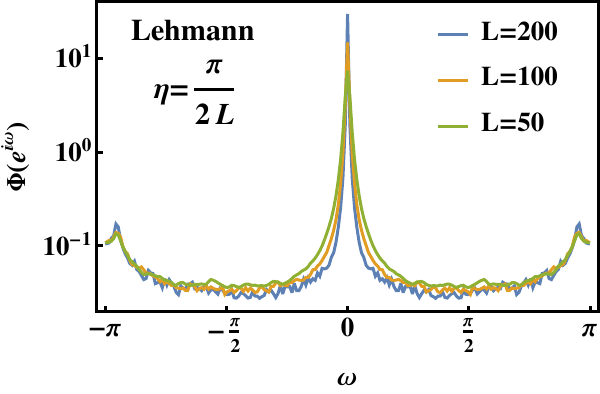}

    \caption{Numerical results for the Lehmann representation of the spectral function for the Anderson model with $h = 10$, averaged over $5000$ realizations. Two different width parameters, $\eta = \pi/L$ and $\eta = \pi/(2L)$, are considered in the left and right panels, respectively, for system sizes $L = 50, 100, 200$. For larger system sizes or smaller $\eta$, the peak at $\omega = 0$ becomes more pronounced, consistent with the emergence of a delta-function in the thermodynamic limit. However, for smaller $\eta$, a larger number of realizations is required to obtain a smooth spectrum.}
    \label{Fig: Lehmann}
\end{figure}

\section{Statistics of the Krylov angles for the Anderson model}
\label{App: Angle statistics}
We present numerical results that reveal the detailed statistical behavior of the Krylov angles in the strong-disorder regime with $h = 5$, and for $T = 1.5$, $T = 30$ (using the same data as in Fig.~\ref{Fig: Anderson-1} and Fig.~\ref{Fig: Anderson-3}). The PDFs of the cosine of the Krylov angles are shown in Fig.~\ref{Fig: statistics}. Specifically, we display the distributions of $\cos\theta_{10}, \cos\theta_{50}, \cos\theta_{100}$ in the top panels and $\cos\theta_{11}, \cos\theta_{51}, \cos\theta_{101}$ in the bottom panels, along with the Gaussian fits. The standard deviation is computed from the Krylov-angle data, analogous to the autocorrelation in \eqref{eq:var}. For large Krylov indices, the distributions become narrower and their mean values approach zero. As shown in Fig.~\ref{Fig: deviation scaling}, the standard deviation $\sigma(\cos\theta_j)$ scales approximately as $j^{-0.2}$ for large $j$.

\begin{figure}[h!]
    \centering
    \includegraphics[width=0.3\textwidth]{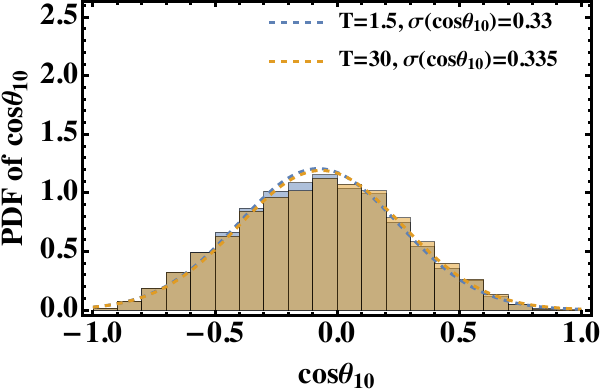}
    \includegraphics[width=0.3\textwidth]{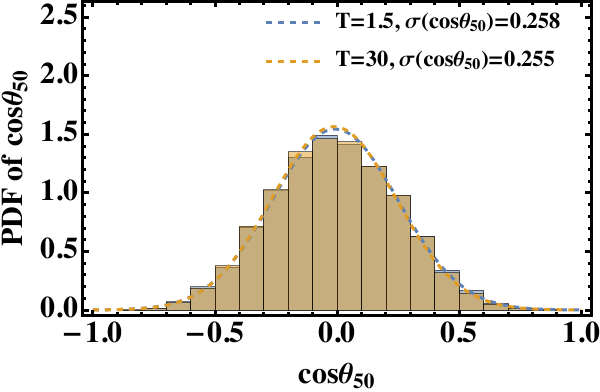}
    \includegraphics[width=0.3\textwidth]{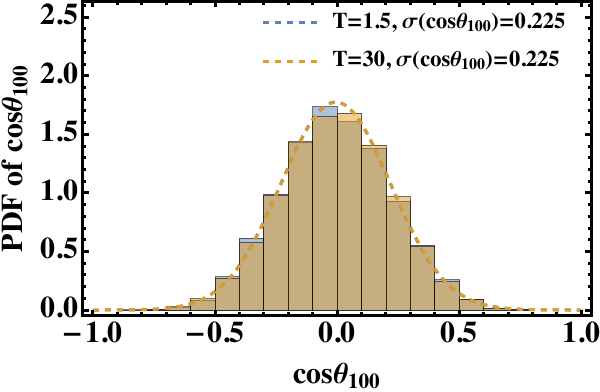}
    \includegraphics[width=0.3\textwidth]{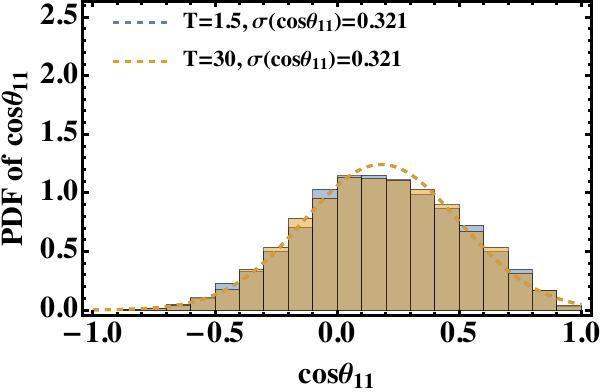}
    \includegraphics[width=0.3\textwidth]{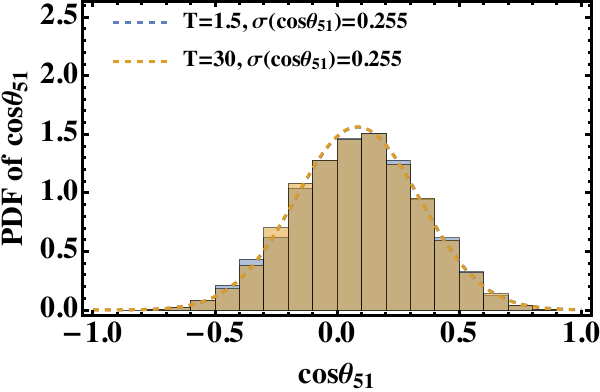}
    \includegraphics[width=0.3\textwidth]{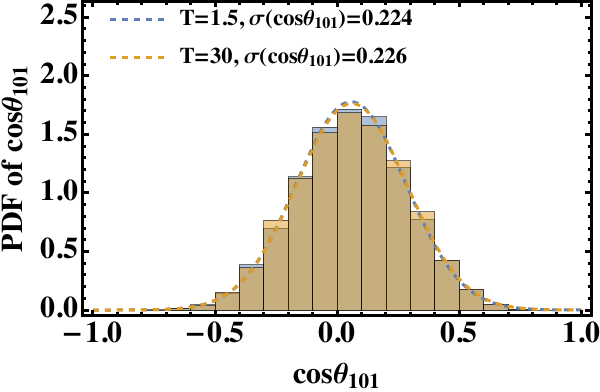}

    \caption{Statistics of the cosine of the Krylov angles for the Anderson model with disorder strength $h = 5$ and system size $L = 200$, for two different stroboscopic time steps, $T = 1.5$ and $T = 30$. The first (second) row shows the PDF of the cosine of even (odd) Krylov angles. For small even (odd) Krylov indices, the distributions are centered to the left (right) and exhibit larger standard deviations. Deeper in Krylov space, the distributions become narrower, are more centered around zero, and are well approximated by a Gaussian distribution (dashed lines in all panels).}
    \label{Fig: statistics}
\end{figure}

\begin{figure}[h!]
    \centering
    \includegraphics[width=0.4\textwidth]{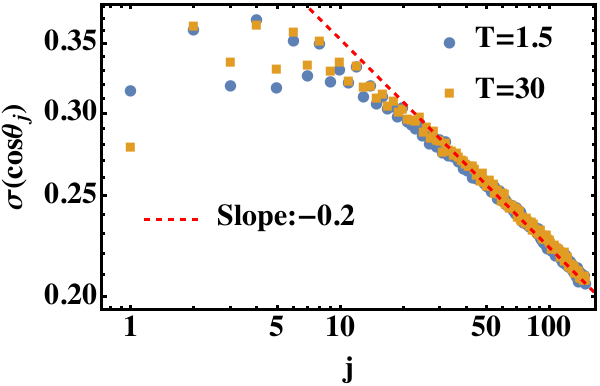}

    \caption{The standard deviation of the cosine of the Krylov angles in the Anderson model with disorder strength $h = 5$ and $L=200$, for two different stroboscopic time steps, $T = 1.5$ and $T = 30$. The data are shown on a log--log scale. The distribution becomes progressively narrower as the Krylov index increases, but the decay is slow and follows a power law with an exponent of approximately $-0.2$.}
    \label{Fig: deviation scaling}
\end{figure}

To further assess whether fluctuations between Krylov angles are correlated, we compute the correlation matrix defined by
\begin{align}
    \rho_{jk} = \frac{\langle( \cos\theta_j - \langle \cos\theta_j \rangle )( \cos\theta_k - \langle \cos\theta_k \rangle )\rangle}{\sigma(\cos\theta_j)\sigma(\cos\theta_k)}.
\end{align}
If the Krylov angles are uncorrelated, $\rho_{jk}$ takes the value $1$ for $j = k$ and remains small for $j \neq k$. The numerical results for the correlation matrix are presented in Fig.~\ref{Fig: angles correlation}, corresponding to the data in Fig.~\ref{Fig: statistics}. The left panel shows $\rho_{jk}$ explicitly for $j = 10$, where the correlation is peaked at $k = 10$ and rapidly decreases for $k \neq 10$. The middle and right panels display color plots of $|\rho_{jk}|$, showing that the diagonal elements are $1$ while the off-diagonal elements remain small for large Krylov indices. This indicates that the Krylov angles are essentially uncorrelated. A similar feature has been reported in Ref.~\cite{peacock2025anderson}, where the Lanczos coefficients become effectively independent deep in the Hamiltonian Krylov space for the Anderson model.

\begin{figure}[h!]
    \centering
    \includegraphics[width=0.3\textwidth]{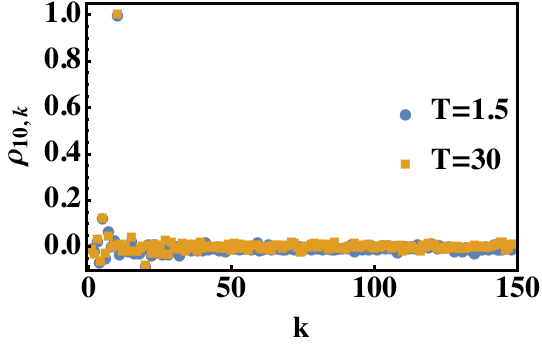}
    \includegraphics[width=0.3\textwidth]{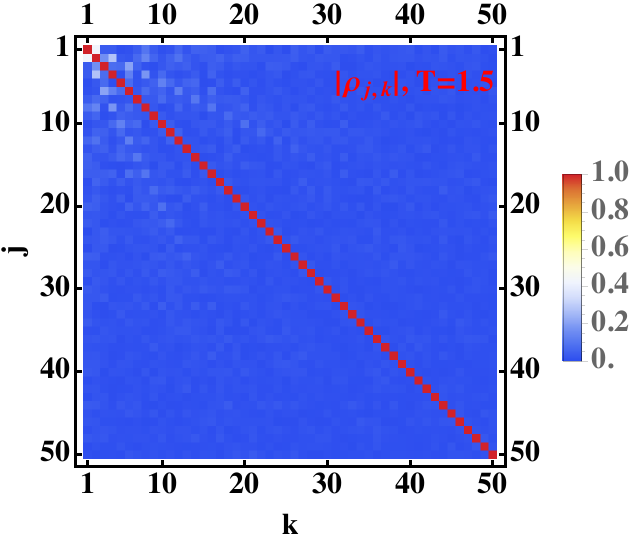}
    \includegraphics[width=0.3\textwidth]{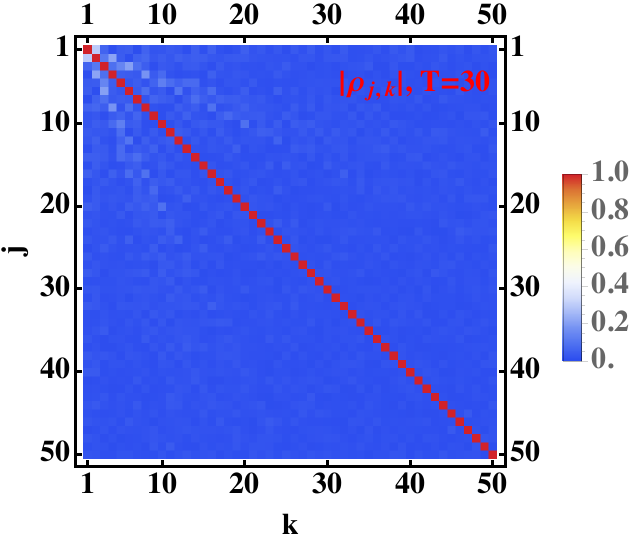}

    \caption{Numerical results for the correlation matrix $\rho_{j,k}$ for the angles in Fig.~\ref{Fig: statistics}. The left panel shows $\rho_{10,k}$ explicitly, indicating that the correlation is peaked at $k = 10$. The color plots of $|\rho_{j,k}|$ are presented in the middle ($T = 1.5$) and right ($T = 30$) panels, demonstrating that the fluctuations of the Krylov angles are essentially uncorrelated.}
    \label{Fig: angles correlation}
\end{figure}

\section{Spectral function and autocorrelation in the delocalized phase of Aubry--André model}
\label{App: AA numerical details}

The results for the approximate spectral function in the delocalized phase and the autocorrelations at the critical point are presented below. Fig.~\ref{Fig: AA delocalized spectrum} shows the approximate spectral functions corresponding to Fig.~\ref{Fig: AA delocalized}. As the system size increases (from left to right panels), the spectrum becomes progressively flatter, consistent with the expected dynamics in the delocalized phase. The peak at $\omega = 0$, which arises from finite-size effects, is also suppressed with increasing system size.

Fig.~\ref{Fig: AA critical autocorrelations} shows the autocorrelation at the critical point for two stroboscopic time steps, $T = 10$ and $T = 100$, corresponding to Fig.~\ref{Fig: AA crital}. For $T = 10$ (left panel), the autocorrelation exhibits larger values at early times compared to later times, indicating that the selected time window does not fully probe the late-time regime. In contrast, for $T = 100$, the autocorrelation displays more uniform fluctuations over time, providing a better characterization of the late-time statistics. This dependence on the stroboscopic time step at the critical point is also reflected in the discrepancy between $\langle A \rangle_t$ and the IPR observed in Fig.~\ref{Fig: AA crital}. 

\begin{figure}[h!]
    \centering
    \includegraphics[width=0.3\textwidth]{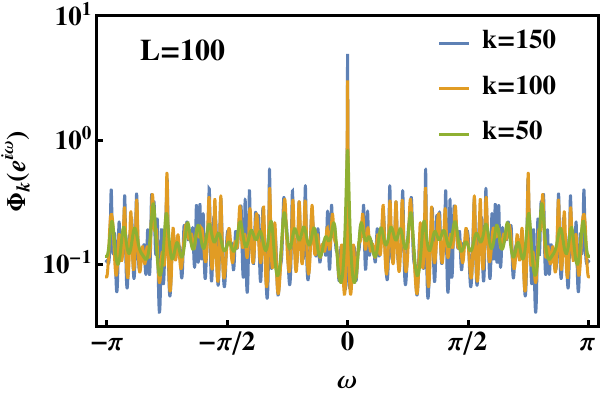}
    \includegraphics[width=0.3\textwidth]{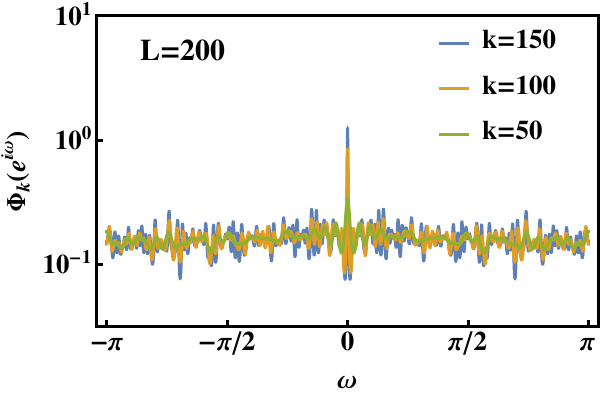}
    \includegraphics[width=0.3\textwidth]{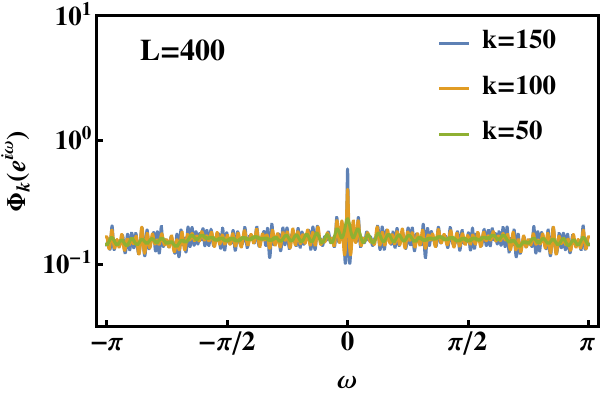}
    \caption{Delocalized phase of the 
    Aubry-Andr\'e model. The approximate Bernstein--Szeg\H{o} spectral functions for $L = 100, 200, 400$ from Fig.~\ref{Fig: AA delocalized} are shown in the left, middle, and right panels, respectively. As the system size increases, the spectrum becomes flatter, consistent with being in the delocalized phase. The peak at $\omega = 0$ reflects the non-decaying component of the autocorrelation arising from finite-size effects.}
    \label{Fig: AA delocalized spectrum}
\end{figure}
\begin{figure}[h!]
    \centering
    \includegraphics[width=0.4\textwidth]{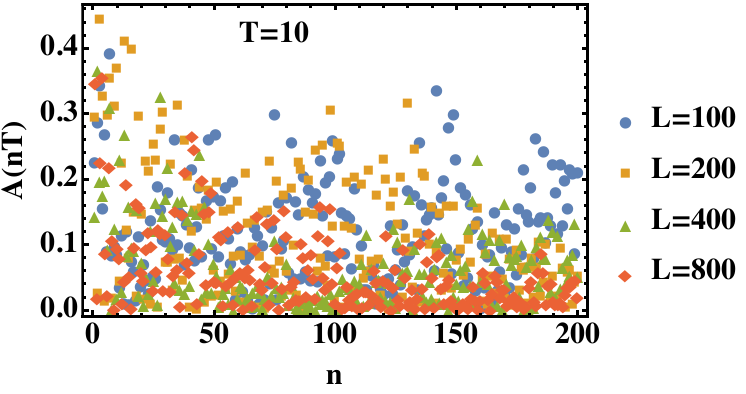}
    \includegraphics[width=0.4\textwidth]{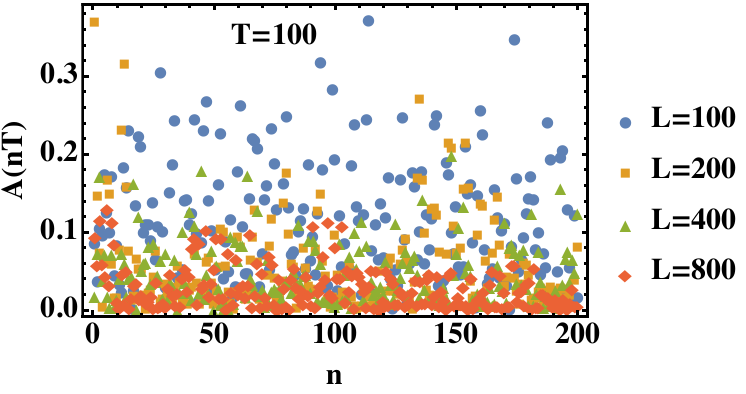}
    \caption{The autocorrelation data at the critical point of the Aubry-Andr\'e model, and used in Fig.~\ref{Fig: AA crital}. For a small stroboscopic time step $T = 10$ (left panel), the autocorrelation exhibits early-time transient behavior, which disappears for a larger time step $T = 100$. This reflects the fact that $\langle A \rangle_t$ requires a sufficiently large $T$ to approach the IPR result in Fig.~\ref{Fig: AA crital}.}
    \label{Fig: AA critical autocorrelations}
\end{figure}

\end{widetext}


%

\end{document}